\documentclass[debug]{rmaa}

\usepackage[frozencache,cachedir=.]{minted}
\usepackage{paralist}
\usepackage{lscape}
\usepackage{multirow}
\usepackage{multicol}
\usepackage{MnSymbol}
\usepackage{graphicx} 
\usepackage{blindtext} 
\usepackage{xcolor} 
\usepackage{minted} 
\usepackage{hyperref}
\usepackage{bm}

\usepackage{psfrag,color}

\usepackage[latin1]{inputenc}

\title{Confirmation and Characterization of Galactic Planetary Nebulae: Insights from a Spectroscopic Study} 

\author{
  D. A. Bele\~{n}o-Molina,\altaffilmark{1} 
  L. Olgu\'{\i}n,\altaffilmark{1}
  L. F. Miranda,\altaffilmark{2}
  M. E. Contreras,\altaffilmark{1}
  and R. V\'{a}zquez\altaffilmark{3}
  }

\altaffiltext{1}{Departamento de investigaci\'{o}n en F\'{\i}sica, Universidad de Sonora (UNISON), Mexico.}

\altaffiltext{2}{Instituto de Astrof\'{\i}sica de Andaluc\'{\i}a (IAA), Consejo Superior de Investigaciones Cient\'{\i}ficas (CSIC), Spain.}

\altaffiltext{3}{Instituto de Astronom\'{\i}a en Ensenada, Universidad Nacional Aut\'{o}noma de M\'{e}xico (UNAM), Mexico.}

\shortauthor{Bele\~{n}o-Molina et al.}
\shorttitle{Insights from a Spectroscopic Study}

\fulladdresses{
\item D. A. Bele\~{n}o-Molina, L. Olgu\'{\i}n and M. E. Contreras: Departamento de Investigaci\'{o}n en F\'{\i}sica, Universidad de Sonora, Blvd. Rosales Esq. L.D. Colosio, Edif. 3H, 83000 Hermosillo,
Son., Mexico (a219230153@unison.mx).
\item L. F. Miranda: Instituto de Astrof\'{\i}sica de Andaluc\'{\i}a -- CSIC, Glorieta de la Astronom\'{\i}a s/n, E--18008 Granada, Spain.
\item R. V\'{a}zquez: Instituto de Astronom\'{\i}a, Universidad Nacional Aut\'{o}noma de M\'{e}xico, Km 103 ctra. Tijuana-Ensenada, 22860 Ensenada, B.C., Mexico.}

\listofauthors{D. A. Bele\~{n}o-Molina, L. Olgu\'{\i}n, L. F. Miranda, M. E. Contreras \& R. V\'{a}zquez}
\indexauthor{D. A. Bele\~{n}o-Molina}
\indexauthor{L. Olgu\'{\i}n}
\indexauthor{L. F. Miranda}
\indexauthor{M. E. Contreras}
\indexauthor{R. V\'{a}zquez}

\abstract{We present a spectroscopic investigation of 25 objects previously reported as possible Planetary Nebulae (PNe) in recent catalogs to obtain their physical properties and to establish their true nature. We found 11 objects showing intense emission lines, 11 where it was not possible to measure $\mathrm{H{\beta}}$, and three where no lines are present. We have used diagnostic diagrams to confirm the true PN nature for eight objects. We obtained elemental abundances for three objects whose values are in agreement with the PNe mean values for our Galaxy. Four objects show [\ion{N}{II}] $\lambda$6583 more intense than $\mathrm{H{\alpha}}$, and for two of them, this can be explained by the presence of shocks in the gas. Finally, we report angular sizes based on  $\mathrm{H{\alpha}}$ and [\ion{O}{III}] $\lambda$5007 emission.}

\resumen{Presentamos un estudio espectrosc\'{o}pico de 25 objetos reportados como posibles Nebulosas Planetarias (NPs) en cat\'{a}logos recientes para obtener sus propiedades f\'{\i}sicas y establecer su verdadera naturaleza. Encontramos 11 objetos que muestran l\'{\i}neas de emisi\'{o}n intensas, 11 donde no fue posible medir $\mathrm{H{\beta}}$ y tres que no muestran l\'{\i}neas. Utilizando diagramas de diagn\'{o}stico, hemos podido confirmar ocho objetos como verdaderas NPs. Para tres de estos objetos hemos determinado abundancias elementales que concuerdan con los valores del conjunto de NPs de nuestra Galaxia. Cuatro objetos muestran la l\'{\i}nea [\ion{N}{II}] $\lambda$6583 m\'{a}s intensa que $\mathrm{H{\alpha}}$ que puede explicarse en dos de ellos con la presencia de choques en el gas. Finalmente, reportamos tama\~{n}os angulares determinados a partir de la emisi\'{o}n en $\mathrm{H{\alpha}}$ y [\ion{O}{III}] $\lambda$5007.}

\addkeyword{Catalogues}
\addkeyword{ISM: Abundances}
\addkeyword{Planetary nebulae: General}
\addkeyword{Techniques: Spectroscopic}

\begin{document}
\maketitle

\section{Introduction}
\label{sec:introduction}

Planetary Nebulae (PNe) are the product of the gas ejection of evolved stars with low and intermediate-mass progenitors ($\sim$0.8--8$\mathrm{M_{\odot}}$). These stars eject the most external layers of their surface at the Asymptotic Giant Branch (AGB) evolutionary stage, with mass loss rates around $10^{-8}\,\mathrm{M_{\odot}\,yr^{-1}}$. At the tip of the AGB phase, mass loss reaches up to $10^{-4}\,\mathrm{M_{\odot}\,yr^{-1}}$, creating an expanding shell around them \citep{2020A&A...636A.123R}. When the uncovered central hot core reaches an effective temperature ($T_\mathrm{eff}$) of approximately 30,000 K \citep{2024AdSpR..74.1366K}, it produces energetic photons capable of exciting and ionizing the previously ejected material, forming a so-called planetary nebula (PN). Subsequent line deexcitation and recombination in the envelope produce the characteristic spectrum associated with these objects \citep{2006agna.book.....O}. Considering a scenario where single and binary stars produce PNe, the predicted Galactic population of these objects is about 46,000 \citep{2006ApJ...650..916M}. However, by 2000, the number of known PNe was approximately 1500 \citep{2001A&A...378..843K}, much smaller than expected. More recently, as of April 2025, the Galactic known PNe population has increased to 3952 objects according to the HASH \citep{2016JPhCS.728c2008P}, an amount larger than the previous findings but still smaller than expected. This discrepancy could be explained by a high dust obscuration in the Galactic plane, by the faintness of old PNe or by the compactness of sources usually identified as stars \citep{2000oepn.book.....K}.\\

During the last 20 years, many sky surveys have searched for these elusive objects, making a golden age of PNe discovery (\citealp{2022PASP..134b2001K, 2022FrASS...9.5287P}). Objects included in these catalogs generally have a small angular extent and some of them are underrepresented by its low surface brightness. Several authors have begun studying these objects (e.g.\@ \citealp{2010ApJ...724..748H, 2012RMxAA..48..223A, 2014A&A...563A..63H, 2016MNRAS.462.1393A, 2023MNRAS.520..773R, 2024MNRAS...527..1481T}) but a considerable number ($\sim$1000) still need an initial study or a deeper analysis of their properties. Since these objects are poorly studied in literature, it is necessary to establish their true nature to understand if they could represent an evolutionary stage not well-studied before and why many of them were difficult to detect. The presence of dust, highly evolved objects, or small size emission regions can be claimed responsible for the latter.\\

Only some of the new objects reported in recent catalogs have been confirmed as true PNe \citep{2014MNRAS.443.3388S}. Furthermore, additional studies are needed to obtain and analyze the physical and chemical properties of those confirmed PNe. In this paper, we present a spectroscopic study of 25 objects reported as potential PNe in catalogs. We derived physical parameters and chemical abundances for some selected objects.

\section{Object selection}
\label{sec:Objselection}

We selected our targets from an exhaustive search in various databases, catalogs, and papers: \citet{1992secg.book.....A}, \citet{2001A&A...378..843K}, IPHAS \citep{2005MNRAS.362..753D, 2009AA...502..113V, 2014MNRAS.443.3388S}, MASH \citep{2006MNRAS.373...79P, 2008MNRAS.384..525M}, \citet{2015LAstr.129a..42F}, HASH \citep{2017IAUS..323..327B, 2016JPhCS.728c2008P}, and PNST\footnote{Planetary Nebula Spectra Trackers (PNST) is a french amateur observational group designed to uncover and confirm new Galactic PNe.} \citep{2022A&A...666A.152L}. Our main selection criterion was objects reported in any catalog as a possible PN, a PN candidate, and confirmed PN with limited or no previous studies.\\ 

Table~\ref{tab1} shows all the known, relevant, and updated information found in the literature about our selected sample of 25 objects. It indicates whether spectra, morphologies, central stars, or distances are available. We added their HASH and SIMBAD statuses for comparison.\\ 

\section{Observations and data reduction}
\label{sec:observations}

Observations were carried out at the Observatorio Astron\'omico Nacional on the Sierra San Pedro M\'{a}rtir (hereafter OAN-SPM), and the Observatorio Astrof\'{\i}sico Guillermo Haro (hereafter OAGH). In both cases we used a 2.1 m class telescope. We had three observing runs: 2016 May 7 to 10 (OAGH), 2022 May 24 to 26 and 2022 July 26 to 28 (OAN-SPM). Table~\ref{tab2} shows the observing log, indicating names, coordinates, exposure times, and corresponding observatories. Observations at OAN-SPM were obtained with a seeing of 3.1{\arcsec} on May and 2.1--2.3{\arcsec} on July, while at OAGH, the seeing was 1.8--2.9{\arcsec}. At OAN-SPM, low-resolution long-slit spectra were obtained using the Boller \& Chivens spectrograph with the SI-2 CCD and a 400 lines mm$^{-1}$ dispersion grating, giving a spectral resolution of $\simeq$ 6 {\AA} (FWHM). In the case of the OAGH, low-resolution long-slit spectra were obtained with the Boller \& Chivens spectrograph, the SITe CCD, and a 150 lines mm$^{-1}$ dispersion grating resulting in a spectral resolution of $\simeq$ 14 {\AA} (FWHM). A slit width of $\simeq$ 2.6{\arcsec} was used in both observatories. The slit was always oriented at a position angle (PA) of 90\arcdeg, except for PN G006.5+08.7, for which a $\mathrm{PA=144\arcdeg}$ was used. Appendix \hyperlink{appendixA}{A} shows the slit positions and extraction zones over-imposed on images of the objects available in HASH. Exposure time for most objects was 1800 s.\\

For flux calibration, a wider slit width of $\simeq$ 13{\arcsec} was used to observe the spectrophotometric standard stars BD+28 4211 and Feige 67. Data reduction was carried out following the IRAF standard procedures for bias, flat, wavelength calibration, background, and cosmic ray corrections.

\section{Results and Discussion}
\label{sec:Results}

From our sample of 25 objects we found: 11 objects with clear emission lines, 11 where few emission lines were detected but it was not possible to measure $\mathrm{H{\beta}}$, and three with no emission lines. Spectra of the 11 objects with clear emission lines are shown in Figure \ref{fig:1}. Appendix \hyperlink{appendixB}{B} shows the spectra for the remaining objects. 

\begin{figure}
\begin{changemargin}{-3cm}{-3cm}
\begin{center}
\includegraphics[scale=0.35]{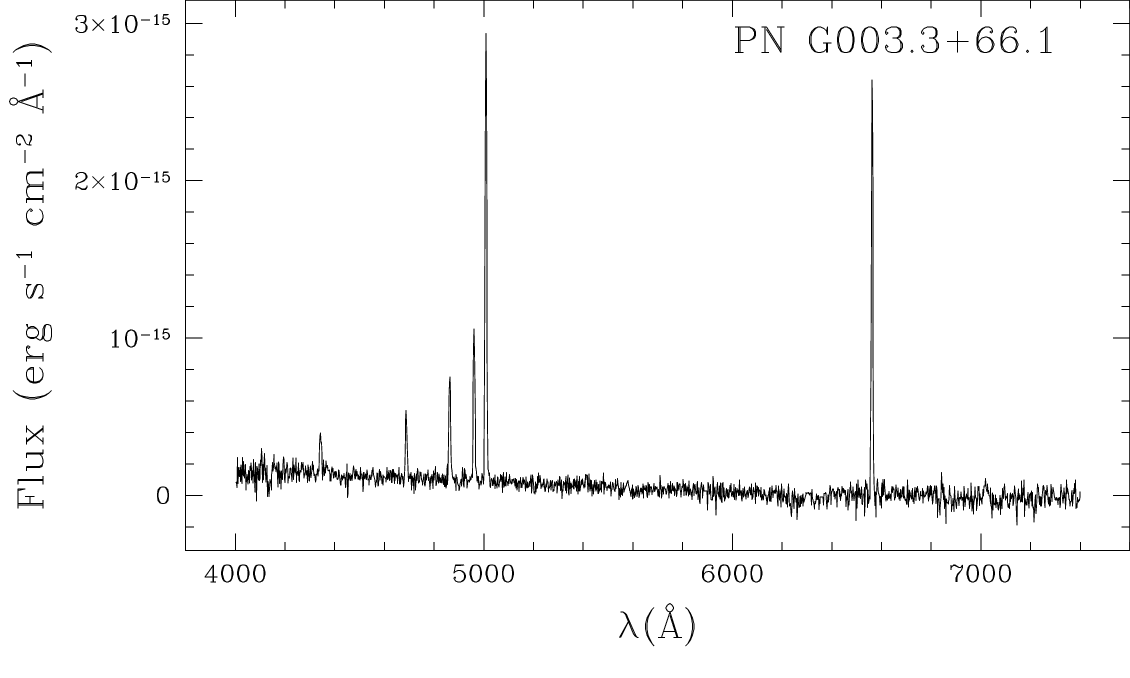} \hspace{0.3 cm} \includegraphics[scale=0.35]{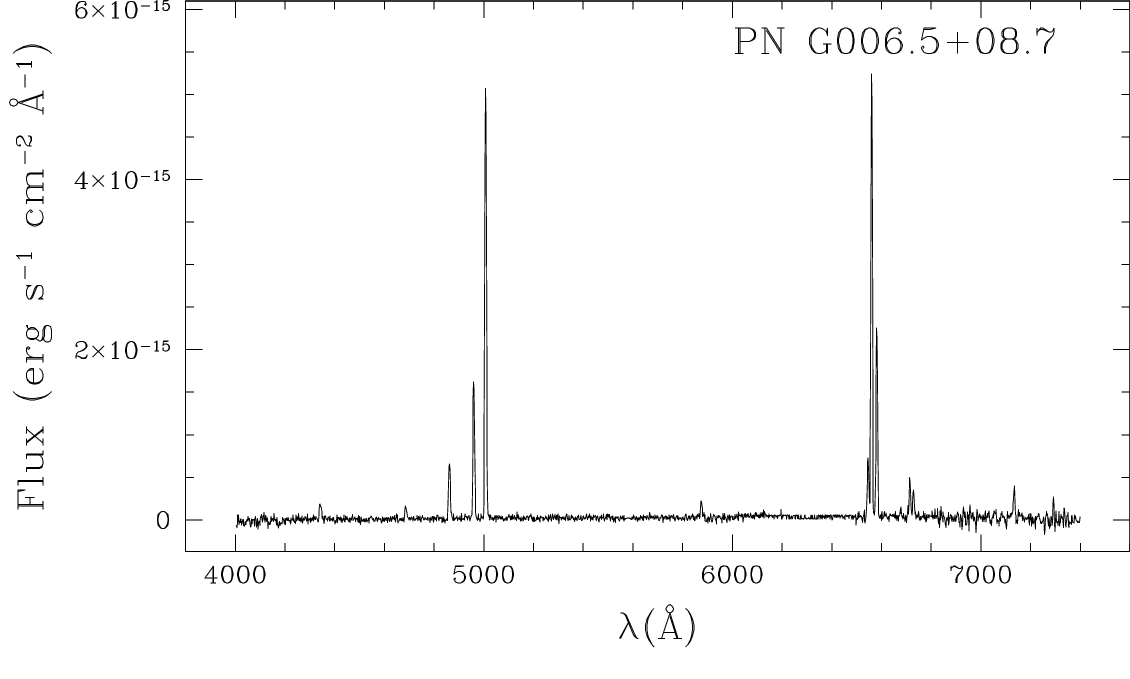}\\ \vspace{0.2 cm}
\includegraphics[scale=0.35]{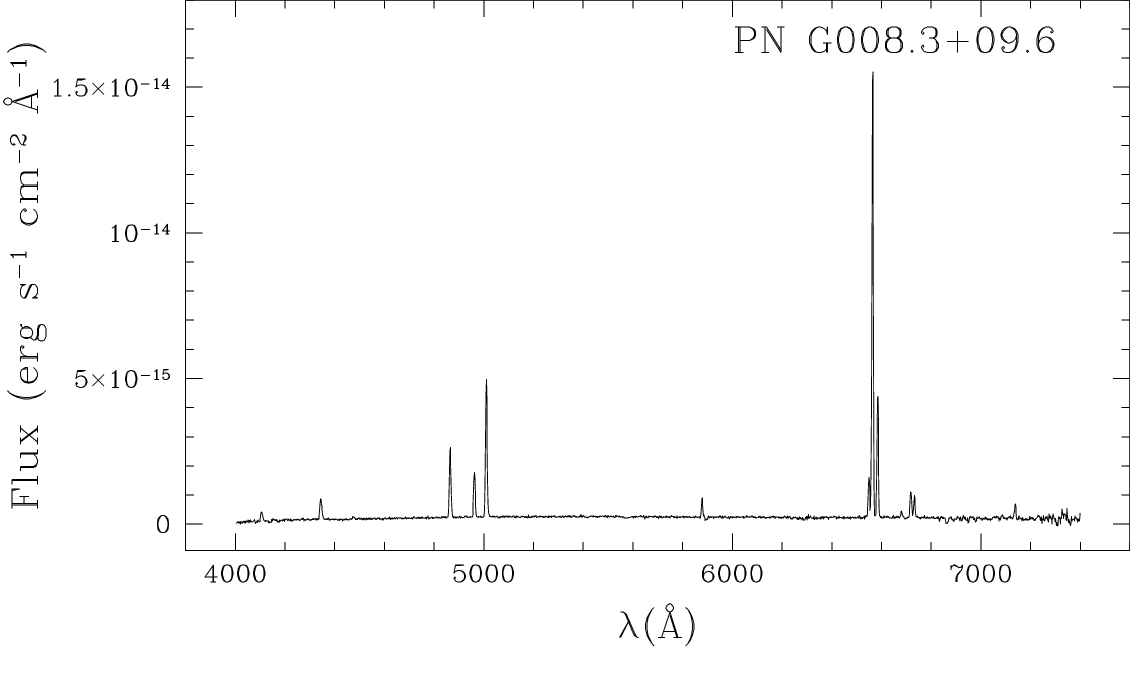} \hspace{0.3 cm} \includegraphics[scale=0.35]{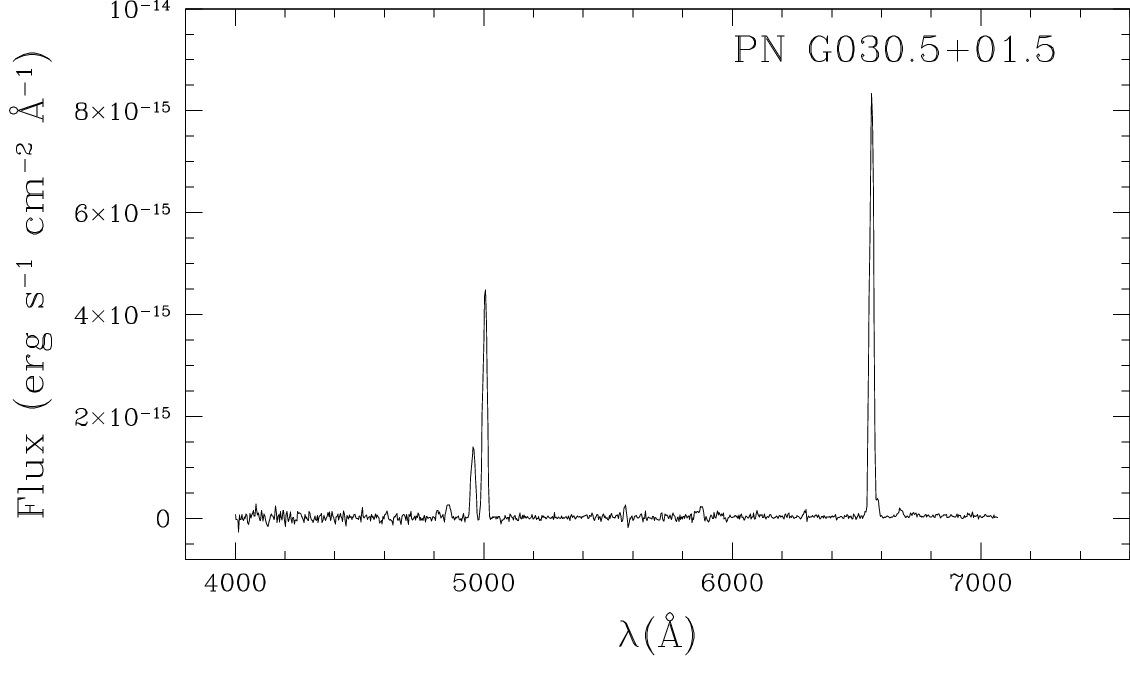}\\ \vspace{0.2 cm}
\includegraphics[scale=0.35]{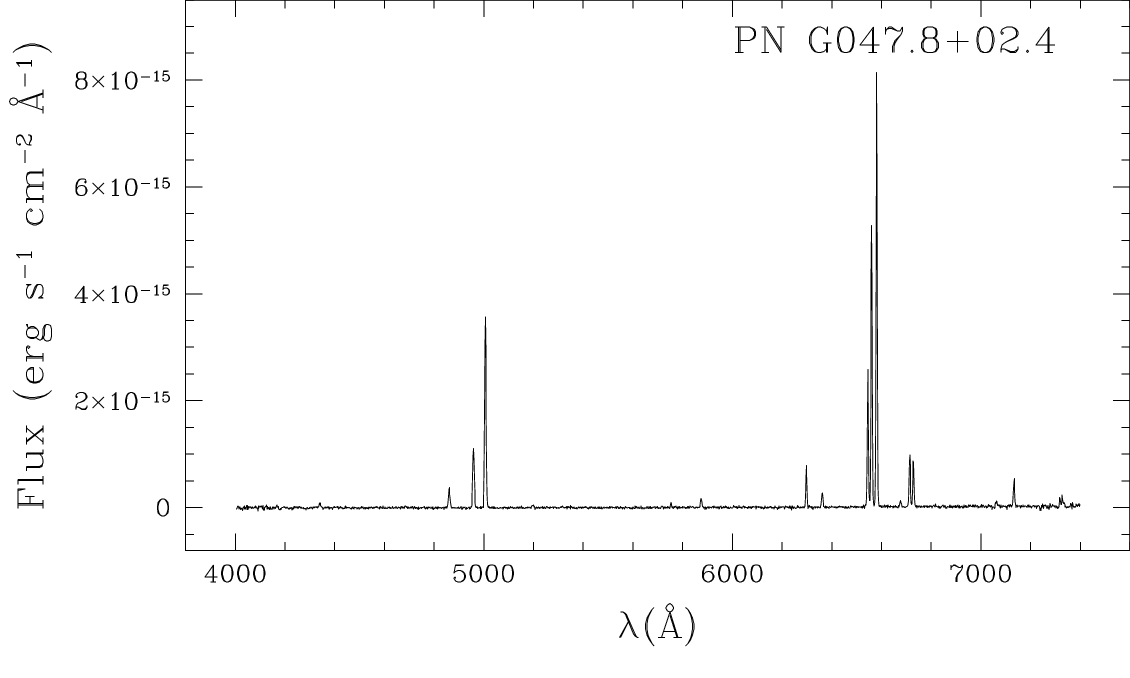} \hspace{0.3 cm} \includegraphics[scale=0.35]{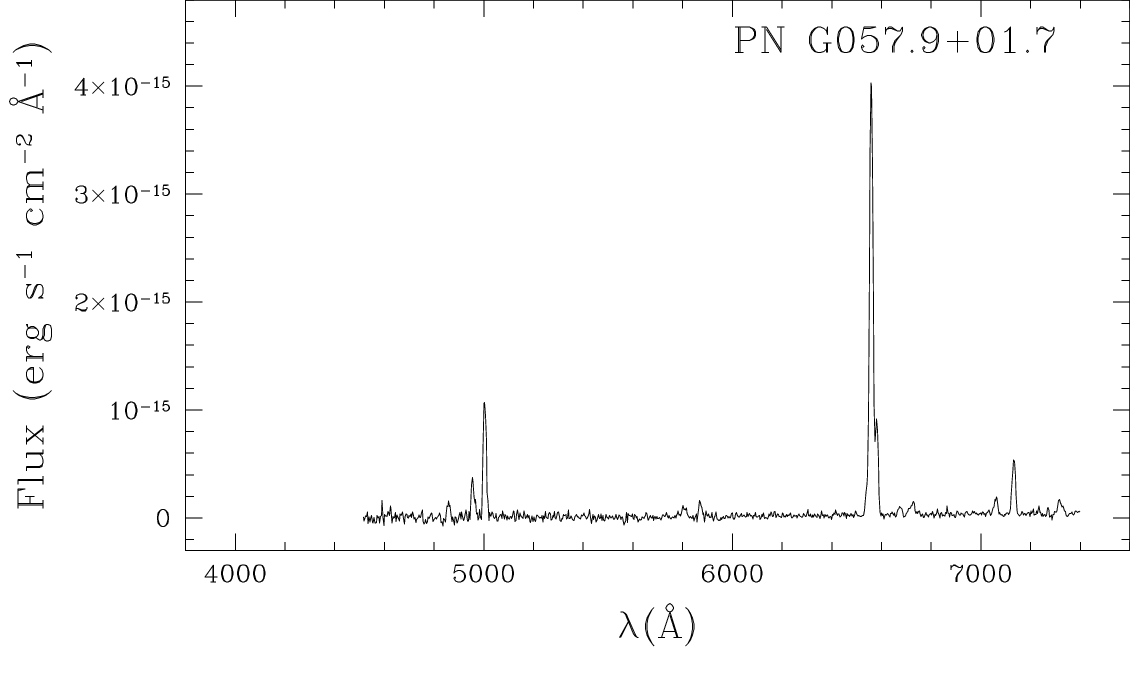}\\
\vspace{0.2 cm}
\includegraphics[scale=0.35]{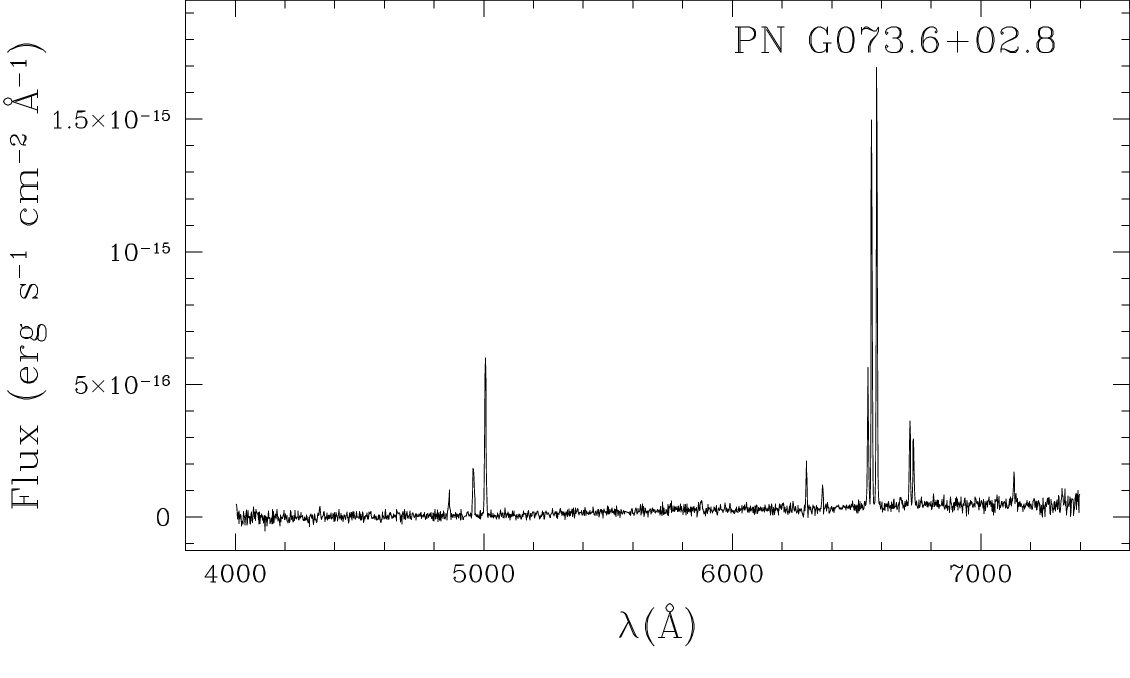} \hspace{0.3 cm} \includegraphics[scale=0.35]{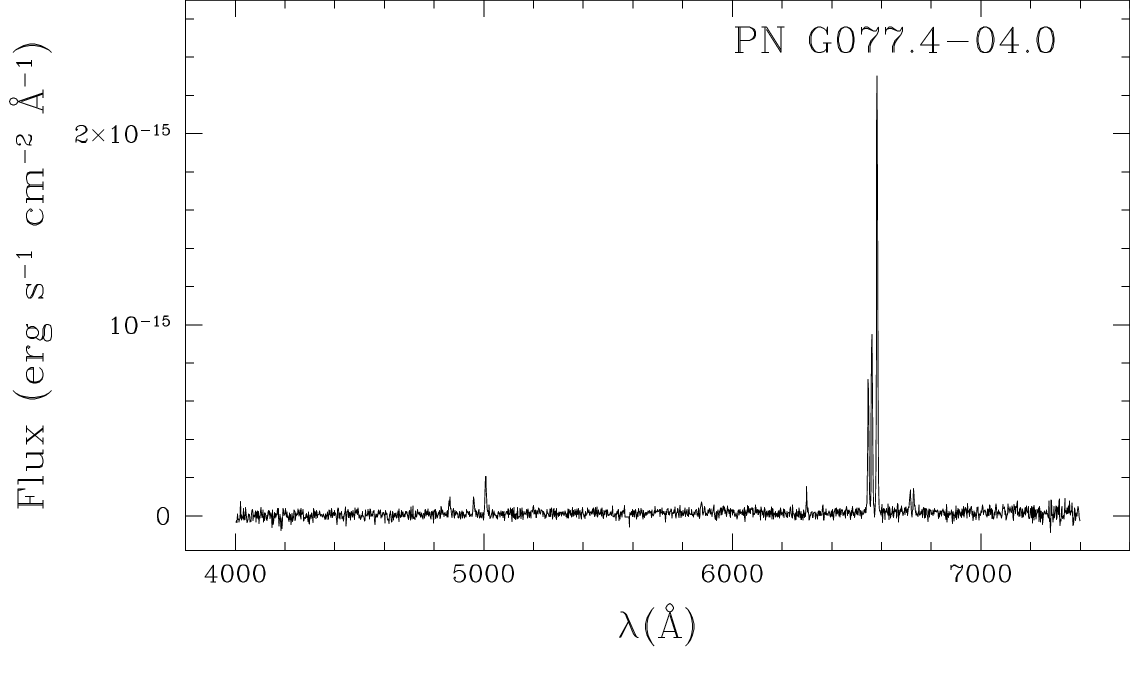}
\end{center}
  \caption{Spectra for 11 selected objects.}\label{fig:1}
\end{changemargin}
\end{figure}

\begin{figure}[H]
\addtocounter{figure}{-1}
\begin{changemargin}{-3cm}{-3cm}
\begin{center}
\includegraphics[scale=0.35]{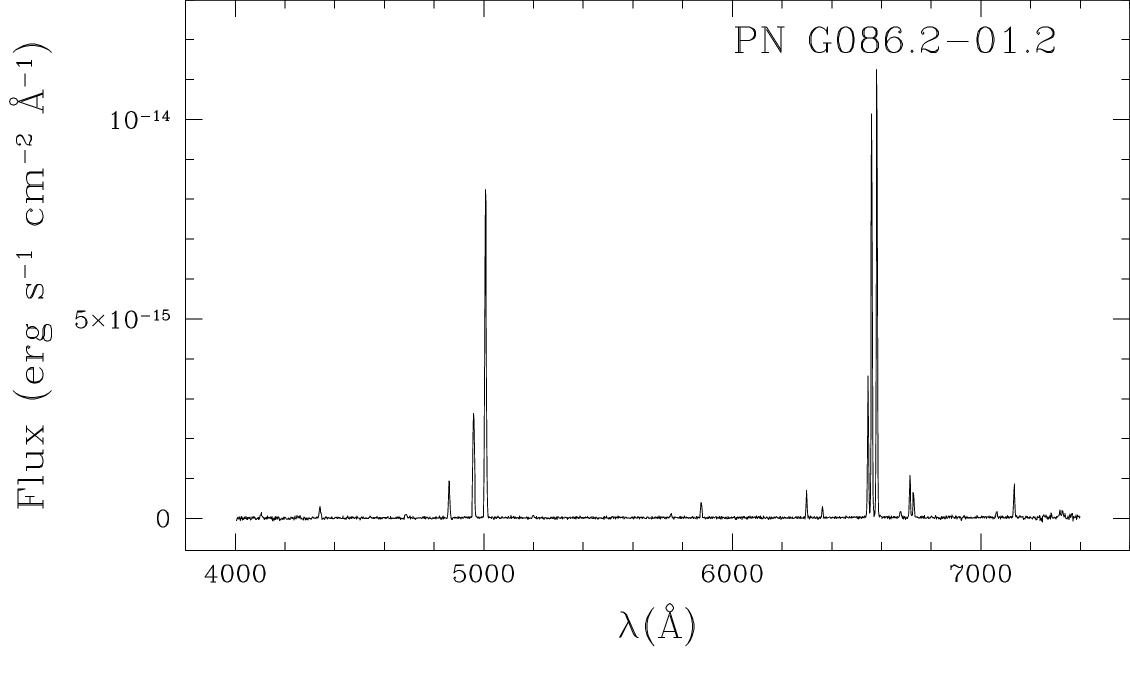} \hspace{0.3 cm} \includegraphics[scale=0.35]{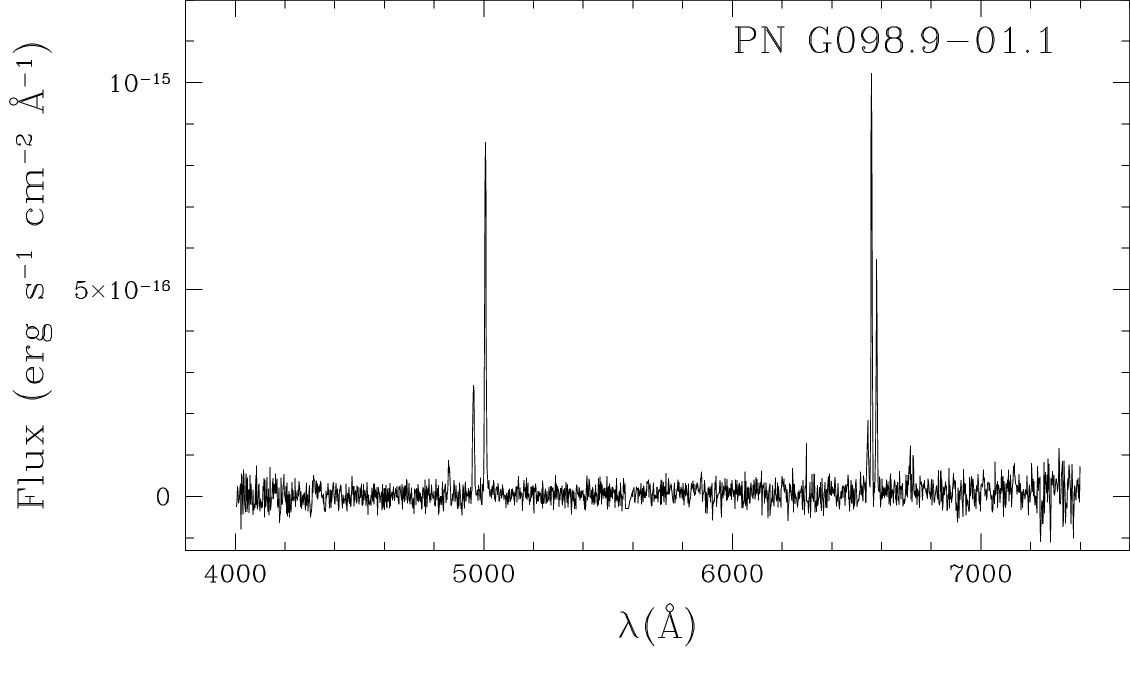}\\
\vspace{0.2 cm}
\includegraphics[scale=0.35]{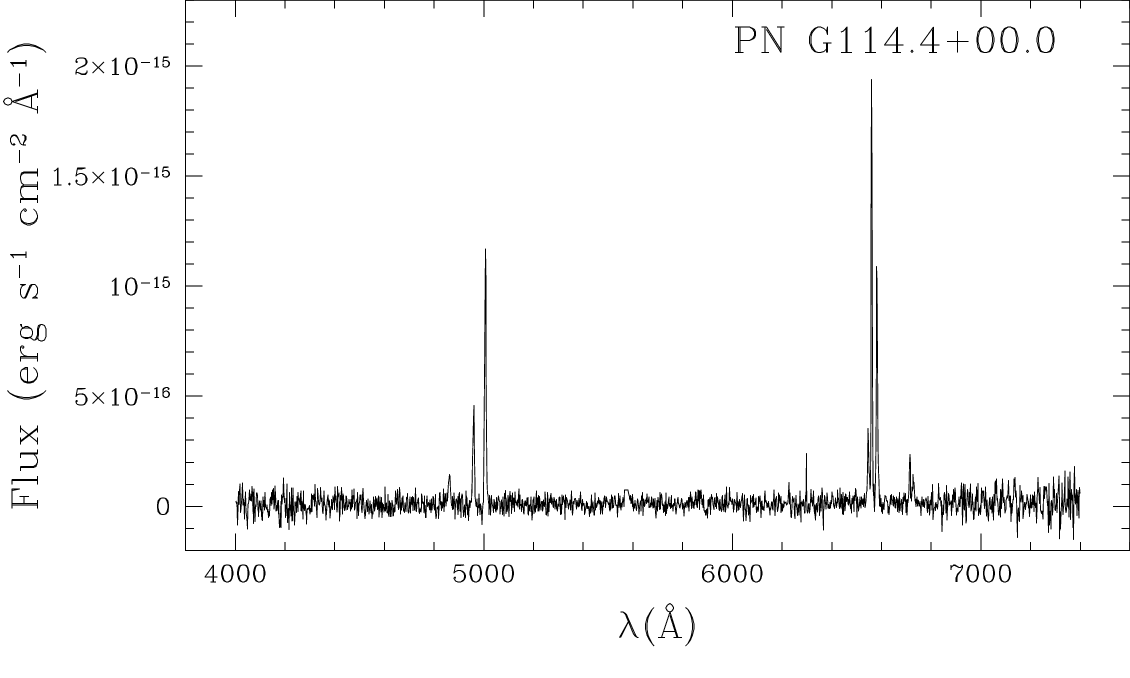}
\end{center}
\caption{(Continued).}
\end{changemargin}
\end{figure}

We used the ANNEB software \citep{2011RMxAC..40..193O} for line identification, to derive the logarithmic extinction coefficient $c(\mathrm{H\beta})$, to obtain intrinsic line intensities and to calculate physical conditions ($T_\mathrm{e}$ and $n_\mathrm{e}$), and ionic and elemental chemical abundances. To obtain ionic abundances, ANNEB uses IRAF's NEBULAR task \citep{1995PASP..107..896S} and for deriving elemental abundances uses ionization correction factors from \citet{1994MNRAS.271..257K}. The uncertainties for all derived quantities were calculated by properly propagating errors in line fluxes, having as a source of error only the number of photons received according the detector gain. Tables~\ref{tab3} and \ref{tab4} show relative line intensity and derived physical parameters for the 11 objects with multiple emission lines. The extinction function of \citet{1989ApJ...345..245C} was adopted to deredden the observed line ratios. $F(\mathrm{H\beta})$ corresponds to the observed $\mathrm{H\beta}$ flux and $SB(\mathrm{H\beta})$ to the surface brightness calculated in the extraction area. Three objects show high extinction with $c(\mathrm{H\beta})>2.0$, which explains the observed weakness in their emission lines.\\

We present electron temperatures from low-ionization species for two PNe derived using the [\ion{N}{II}] $(\lambda{6548}+\lambda{6583})/\lambda{5755}$ line ratio. The [\ion{O}{III}] $\mathrm{\lambda4363}$ line was not detected in our sample, preventing the determination of electron temperature using medium-ionization species. Also, we present electron densities from low-ionization species for eight PNe derived using the [\ion{S}{II}] $\lambda{6716}/\lambda{6731}$ line ratio. The electron densities in the objects where it could be measured are low, which could indicate that they are possible already evolved PNe. Another possible explanation could be PNe expanding at typical rates but where the ejected mass was very low, implying low-mass progenitor stars.\\

\begin{figure}[h!]
  \begin{center}
  \includegraphics[scale=0.58]{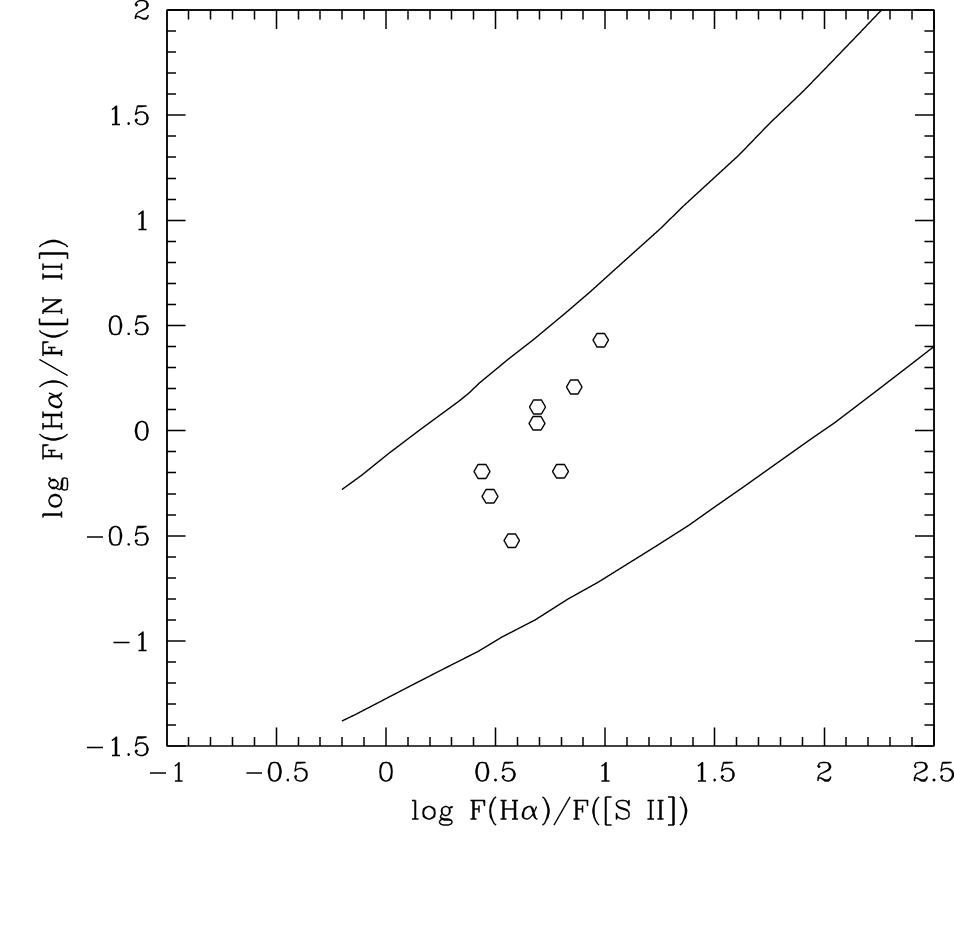}
  \end{center}
  \vspace{-1.5cm}
  \caption{A $\log F(\mathrm{H{\alpha}})/F([\mathrm{\ion{N}{II}}])$ versus $\log F(\mathrm{H{\alpha}})/F([\mathrm{\ion{S}{II}}])$ diagnostic diagram. $F([\mathrm{\ion{N}{II}}])$ refers to the sum of the flux of $\lambda6548$ and $\lambda6583$ nitrogen lines. $F([\mathrm{\ion{S}{II}}])$ refers to the sum of $\lambda6716$ and $\lambda6731$ sulfur lines. All eight objects lie within the PNe region, as indicated by the empirical boundaries from \citet{2010PASA...27..129F}. Uncertainties bars are the same size or smaller than the symbols.}
  \label{fig:2}
\end{figure}

Figure \ref{fig:2} show a diagnostic diagram (DD) following \citet{2010PASA...27..129F}. Eight of our objects are located inside the True PNe region.

Table~\ref{tab5} shows the mean ionic abundances for the 11 selected objects, calculated from the collisionally excited and the optical recombination lines we found. In cases where we could obtain $T_\mathrm{e}$ and $n_\mathrm{e}$, these values were used in the calculations. Where it was not possible, we adopted  $T_\mathrm{e}=10,000$ K and $n_\mathrm{e}=100\, \mathrm{cm^{-3}}$. Table~\ref{tab6} shows elemental abundances in those three objects for which it was possible to derive them. We added some abundance references for the Sun \citep{2021AA...653A.141A}, Galactic disk PNe (southern in \citealp{1994MNRAS.271..257K}, northern in \citealp{2006ApJ...651..898S}) and Galactic bulge PNe \citep{2024MNRAS.527.6363T}. Values in our data are supersolar, except for $\mathrm{O/H}$ and $\mathrm{S/H}$ abundances in PN G086.2$-$01.2 (PN Ra 5) that are sub-solar. Such a low $\mathrm{S/H}$ abundance could be explained by the ``sulphur anomaly'' \citep{2012ApJ...749...61H}. We found congruent elemental abundances with the average values for PNe in our Galaxy. Following classification criteria from \citet{2007A&A...475..217Q}, PN G047.8+02.4 can be classified as a Peimbert type IIa and PN G086.2$-$01.2 as a Peimbert type I. These results are in agreement with the original criteria used by \citet{1983IAUS..103..233P}. Type I PNe are related to massive progenitor stars ($\sim$2.4--8.0 $\mathrm{M_{\odot}}$), while type II are related to less massive progenitors ($\sim$1.2--2.4 $\mathrm{M_{\odot}}$, \citealp{2007A&A...475..217Q}). Following equations developed by \citet{2010A&A...512A..19M}, we have estimated the central star mass ($m_\mathrm{CS}$) and the stellar mass in the main-sequence ($m_\mathrm{MS}$) obtaining $m_\mathrm{CS}=0.6\,\mathrm{M_{\odot}}$ and  $m_\mathrm{MS}=1.7 \,\mathrm{M_{\odot}}$ for PN G047.8+02.4 and $m_\mathrm{CS}=0.7\,\mathrm{M_{\odot}}$ and $m_\mathrm{MS}=2.3\,\mathrm{M_{\odot}}$ for PN G086.2$-$01.2. Following the methodology proposed by \citet{2010PASA...27..187R} and using our dereddened line intensity ratios, we could compute the excitation class (EC) based on the presence or absence of \ion{He}{II} $\mathrm{\lambda4686}$ in the nebular spectrum, which defines the EC on a PN. When a PN has $EC < 5$, it is considered a low EC; if $EC \geq 5$, it is deemed medium to high EC. We found that PN G047.8+02.4 and PN G086.2$-$01.2 are classified in the medium to high EC with values of 5.7 and 6.9, respectively. In addition, adopting the criterion of $F([\ion{N}{II}]\lambda6583)/F(\mathrm{H{\alpha}})\geq1$ as the condition for optically thick PNe \citep{1989ApJ...345..871K}, PN G047.8+02.4 and PN G086.2$-$01.2 are optically thick. Therefore, we can use the empirical relationship between $EC$ and $T_\mathrm{eff}$, defined by \citet{2010PASA...27..187R}, to estimate the central star (CS) temperature. In this way, we have found $T_\mathrm{eff}\,{\simeq}\, 1.1\times10^{5}$ K for PN G047.8+02.4 and $T_\mathrm{eff}\,{\simeq}\, 1.4\times10^{5}$ K for PN G086.2$-$01.2.\\  

\begin{figure}
\begin{center}
  \includegraphics[scale=0.56]{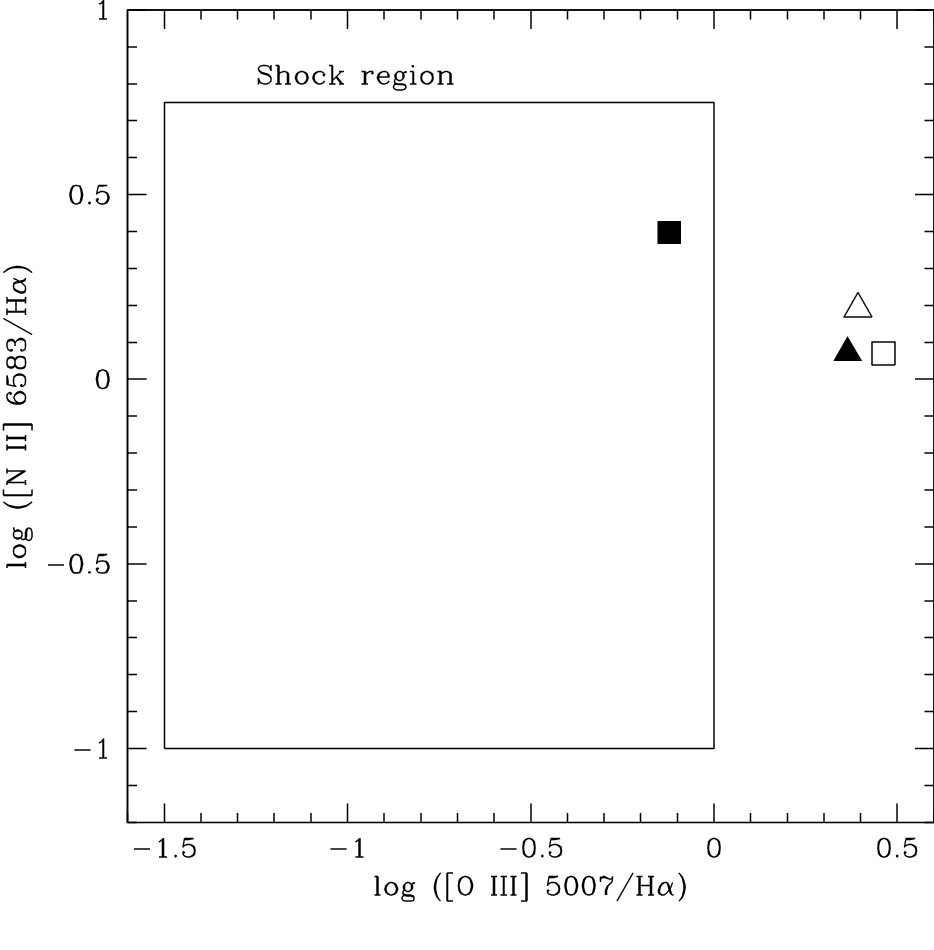}\\
  
  \includegraphics[scale=0.565]{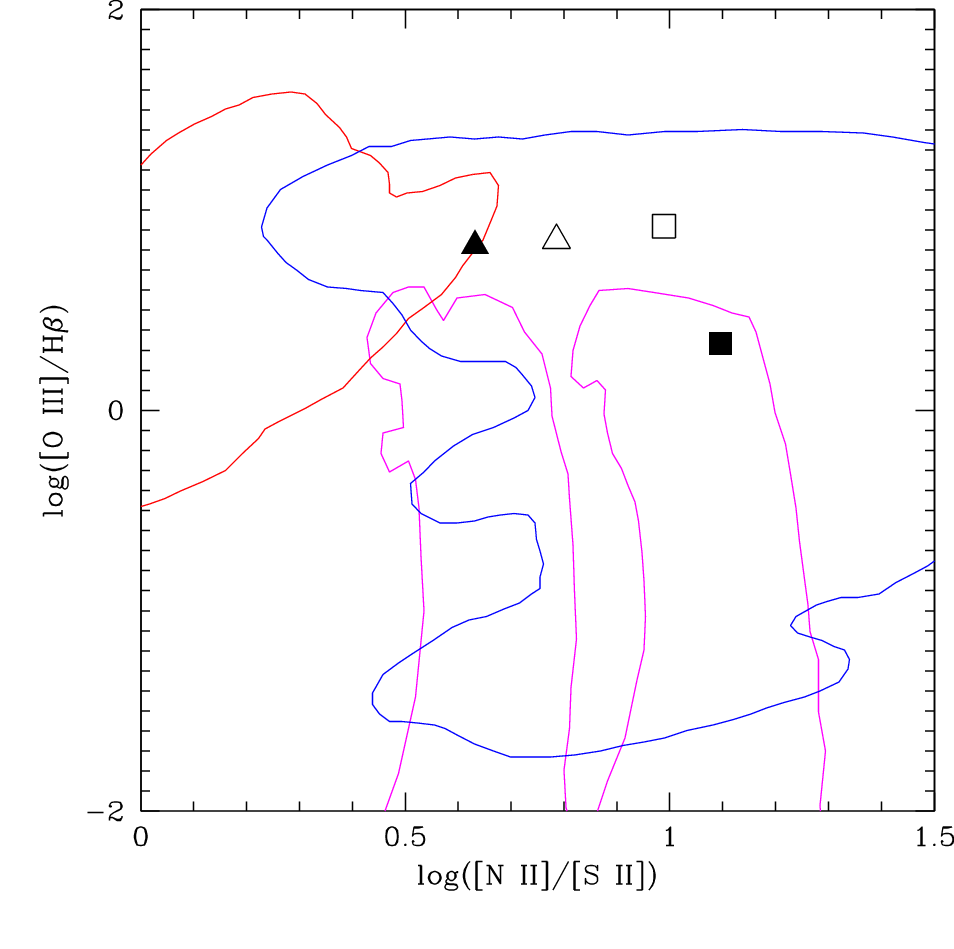}
  \end{center}
  
  \vspace{-1cm}
  \caption{Diagnostic diagrams to distinguish shock-excited and photoionised regions. Top: proposed by \citet{2008A&A...489.1141R} with the shock region from \citet{2017PASA...34...36A}. Bottom: proposed by \citet{2023MNRAS.525.1998M}. Photoionization models are in a blue contour, low-velocity shock models with magenta and high velocity shock models with red. Symbols: $\triangle$ PN G047.8+02.4, $\blacktriangle$ PN G073.6+02.8, $\square$ PN G086.2$-$01.2, and $\blacksquare$ PN G077.4$-$04.0. Uncertainties are the same size or smaller than the symbols.}\label{fig:3}
\end{figure}

Four low ionization objects were found (PN G047.8+02.4, PN G073.6+02.8, PN G077.4$-$04.0, and PN G086.2$-$01.2). For these objects the [\ion{N}{II}] $\lambda$6583 line is much more intense than $\mathrm{H{\alpha}}$, especially the case of PN G077.4$-$04.0 (PN Ra 4) where [\ion{N}{II}] $\lambda$6583 is more than twice as intense as $\mathrm{H{\alpha}}$. This phenomenon could be related to shock processes in the nebula as an excitation mechanism \citep{2016MNRAS.455..930A}. To verify this possibility, we have used \citet{2008A&A...489.1141R}, \citet{2017PASA...34...36A} and \citet{2023MNRAS.525.1998M} diagnostic diagrams, as shown in Figure \ref{fig:3}. We found that PN G077.4$-$04.0 is inside a shock region and the excess of the [\ion{N}{II}] $\lambda$6583 line could be explained by photoionization and low-velocity shock models. In addition, PN G073.6+02.8 is very close to the contour of the high-speed shock model. This may indicate that a weaker shock is taking place in the nebula. [\ion{N}{II}] $\lambda$6583 excess for the remaining two objects could be explained by photoionization models. The central stars of each object in this subgroup must have a low effective temperature.\\ 

Table~\ref{tab7} includes 11 objects where measuring a reliable $\mathrm{H{\beta}}$ was not possible. However, other emission lines were detected: two objects with several lines and nine with at least evident $\mathrm{H{\alpha}}$ emission. $F(\mathrm{H\alpha})$ and $SB(\mathrm{H\alpha})$ are defined as in Tables~\ref{tab3} and \ref{tab4} but for $\mathrm{H\alpha}$. $c(\mathrm{H\beta})_\mathrm{limit}$ is the lower limit of $c(\mathrm{H\beta})$ derived using an estimation of the upper limit of the $\mathrm{H{\beta}}$ flux. This limit was estimated by adding synthetic $\mathrm{H\beta}$ line profiles of different intensities to our original spectrum and finally choosing a flux where the line was clearly seen in the resulting spectrum. The selected upper limit was typically three times the noise level. We have discarded the four noisiest spectra in the $c(\mathrm{H\beta})_\mathrm{limit}$ calculations because the detected $\mathrm{H{\alpha}}$ had a very low signal to noise (S/N).

\newpage

Table 8 shows angular sizes of the PNe as derived from the $\mathrm{H{\alpha}}$ and [\ion{O}{III}] $\lambda$5007 emission lines in the 2D long-slit spectra. Only the region with a flux above 1\% of the maximum intensity of the line was considered. Some of our values are similar to the major diameters reported by different authors, others are new to the literature. All sizes are affected by seeing at the moment of observation. In addition, in a few cases, the PA of the slit was misaligned with the main axis of the source, which could explain the size differences with values reported in the literature. The most significant difference we found was on PN G003.3+66.1. From our sample, some objects have been observed before by different researchers, as evidenced in Table~\ref{tab1}, but no deep studies have appeared in the literature yet. We want to point out that despite the existence of some uploaded spectra in the HASH database, we decided not to use them because presumably are being analyzed by the original authors. Therefore, we have used our spectra in this work. A few objects with spectroscopic or related studies reported in the literature are discussed in the next section. Additional literature information for other objects is available on Appendix \hyperlink{appendixC}{C}.

\section{Notes on individual objects}
\label{sec:individual}

\textit{PN G008.3+09.6:} \citet{2006MNRAS.367.1551B} named this object as PTB 26. These authors reported an optical spectrum where ten emission lines were detected. They derived $c(\mathrm{H{\beta}})=0.62$ and an electron density $n_\mathrm{e}<$ 70 cm$^{-3}$. We found 14 emission line fluxes, $c(\mathrm{H{\beta}})=0.96$ and an electron density of $n_\mathrm{e}=$ 147 cm$^{-3}$. Also, we report mean ionic abundances.\\

\textit{PN G030.5+01.5:} HASH status is True PN, while SIMBAD reports it as an emission-line star based on \citet{1989AJ.....98.1354R}. In our work, we found the expected characteristics of a PN (see emission lines, abundances and spectrum in the text) but it does not appear on Figure \ref{fig:2}, because the [\ion{S}{II}] ($\lambda\lambda6716, 6731$) emission in our spectrum was found not reliable.\\ 

\textit{PN G040.1+03.2:} We found probable \ion{C}{IV} broad line at 5806 {\AA}. The object has also been observed by \citet{2014MNRAS.443.3388S} and the broad emission is clear too. This could be related to a PN with a Wolf$-$Rayet CS, displaying a blended \ion{C}{IV} doublet (at 5801 and 5812 {\AA}) appearing as a broad feature, due to the rapid expansion of the hot stellar wind \citep{2011MNRAS.414.2812D}.\\ 

\textit{PN G041.5+01.7:} Named 1858+0821 by \citet{1996A&AS..118..243V}. The strength of [\ion{N}{II}]$\mathrm{\lambda6548}$ and $\mathrm{\lambda6583}$ (relative to $\mathrm{H{\alpha}}$=100) that they reported from a very low S/N spectrum were 24.4 and 66.4, respectively. In our spectrum we only found $\mathrm{H{\alpha}}$ and [\ion{N}{II}] $\mathrm{\lambda6583}$. The [\ion{N}{II}]$\mathrm{\lambda6548}$ line is blended with $\mathrm{H{\alpha}}$. To correct the $\mathrm{H{\alpha}}$ emission by the $\mathrm{\lambda6548}$ contribution, we estimated $\mathrm{\lambda6548}$ using a theoretical ratio [\ion{N}{II}]$I_{6583}/I_{6548}=2.96$ \citep{2019AJ....158..145U}. The final strengths of the [\ion{N}{II}] lines relative to $\mathrm{H{\alpha}}$ are 16 for $\mathrm{\lambda6548}$ and 47 for $\mathrm{\lambda6583}$. Differences found in the ratios can be explained by the very low S/N of both spectra. Deeper spectrum is needed in order to better understand the nature of this object. However, \citet{2009A&A...501..539U} reported a 6cm radio flux density at 5GHz and associated the source with an object of PN nature.\\

\section{Conclusions}
\label{sec:conclusions}

We report a low-resolution spectroscopic investigation of 25 objects suspected to be PNe. Using diagnostic diagrams, we confirmed eight objects as true PNe, while the remaining objects did not exhibit enough emission lines for reliable classification. The true PNe include PN G086.2$-$01.2 that was previously classified as doubtful and reported as a candidate status in SIMBAD. We have calculated electron temperatures and densities for two and eight PNe respectively, where the S/N in the nebula spectra was appropriate. Ionic abundances are reported for 11 objects and elemental abundances for three objects. Elemental abundances are entirely consistent with Galactic PNe (\citealp{1994MNRAS.271..257K}, \citealp{2006ApJ...651..898S} and \citealp{2024MNRAS.527.6363T}). We derive masses for the central stars of PN G047.8+02.4 and PN G086.2$-$01.2 of 0.6 and 0.7 M$_\odot$, respectively. Peimbert types are also reported for these two objects. The large [\ion{N}{II}] $\lambda$6583/$\mathrm{H{\alpha}}$ line intensity ratio in PN G077.4$-$04.0 and PN G073.6+02.8 of 2.5 and 1.2, respectively, suggests a contribution of shock excitation, as indicated by shock models. A few emission lines were detected for 11 objects for which we were not able to measure $\mathrm{H{\beta}}$. Deeper spectra are needed to confirm their true nature. We have measured $\mathrm{H{\alpha}}$ and [\ion{O}{III}] $\lambda$5007 angular sizes for our sample where possible. Three remaining objects are misclassified as possible PNe, being PN G339.4+29.7 a late-type star, and PN G358.5+09.1 and PN G358.4+08.9 two galaxies.

\section{Acknowledgements}
\label{sec:Acknowledgements}

We thank the anonymous referee for providing comments and suggestions that have improved the manuscript. Based upon observations carried out at the Observatorio Astron\'{o}mico Nacional on the Sierra San Pedro M\'{a}rtir (OAN-SPM), Baja California, M\'{e}xico. We thank the daytime and night support staff at the OAN-SPM for facilitating and helping obtain our observations. This publication is also based on data collected at the Observatorio Astrof\'{\i}sico Guillermo Haro (OAGH), Cananea, M\'{e}xico, operated by the Instituto Nacional de Astrof\'{\i}sica, \'{O}ptica y Electr\'{o}nica (INAOE). Funding for the OAGH has been provided by the Consejo Nacional de Humanidades, Ciencias y Tecnolog\'{\i}as (CONAHCYT). DABM acknowledges support from CONAHCYT (Mexico) grant 806246. LFM acknowledges support from grants PID2020-114461GB-I00, PID2023-146295NB-I00, and CEX2021-001131-S, funded by MCIN/AEI/10.13039/501100011033. RV acknowledges support from UNAM-DGAPA-PAPIIT grant IN103125. This research used the SIMBAD database, operated at CDS, Strasbourg, France, and the HASH PN database in hashpn.space.

\newpage

\begin{landscape}
\begin{table}  
\begin{changemargin}{-1cm}{-1cm}
\begin{center} 
\vskip -3.5cm
\caption{Literature information of the observed objects} 
\label{tab1}
\vskip 0.1cm
\begin{tabular}{lccccccc} \hline                         
Name & HASH optical spectra$^a$ & Morph.$^b$ & Central Star$^c$ & $d_\mathrm{geo}$ (pc)$^d$ & $d_\mathrm{phot}$ (pc)$^d$ & HASH$^e$ & SIMBAD$^f$\\
\hline 
PN G003.3+66.1 & ... & R &  1227151802340968832 & $729_{-126}^{+176}$ & $464_{-18}^{+32}$ & P & PN\\
PN G339.4+29.7 & ... & ... & ... & ... & ... & NPN & PNC\\
PN G358.5+09.1 & ... & ...  & ... & ... & ... & PG & PNC\\
PN G358.4+08.9 & ... & ...  & ... & ... & ... & PG & PNC\\
PN G006.5+08.7 & ... & Ea$\ast$ & 4122534632633972992 & $3327_{-602}^{+853}$ & $3374_{-1058}^{+2278}$ & T & PN\\
PN G008.3+09.6 & SO, 2002, 0.3m, 1$\times$3600s & B$\ast$ & 4124562102059801216 & $8402_{-1735}^{+2263}$ & $7996_{-1495}^{+1518}$ & T & PN\\
PN G031.3+02.0 & ... & S & ... &  ... & ... & P & S\\
PN G030.5+01.5 & SSO, 2008, 2.3m, 1$\times$300s & E & 4260148618930827008 & ... & ... & T & ELS\\
PN G040.1+03.2 & SAAO, 2011, 1.9m, 1$\times$1200s & Eam$\dagger$ & 4310230750781889920 & $7961_{-3893}^{+3458}$ & $4786_{-1662}^{+1331}$ & T & PN\\
PN G038.4+02.2 & KPNO, 2009, 2.1m, 1$\times$1800s & R$\dagger$ & ... & ... & ... & T & PN\\
PN G040.6+01.5 & SAAO, 2011, 1.9m, 1$\times$1200s & B/S$\ddagger$ & ... & ... & ... & T & PN\\
PN G041.5+01.7 & KPNO, 1993, 2.1m, 1$\times$900s & S$\dagger$ & 4310136536377607040 & ... & ... & L & PN\\
PN G047.8+02.4 & KPNO, 2009, 2.1m, 1$\times$1800s & E$\dagger$ & ... & ... & ... & T & PN\\
PN G050.0+01.0 & ... & E & 4320871725541398400 & ... & ... & T & PNC\\
PN G051.8+01.7 & ... & S & 4514907211117285504 & ... & ... & P & S\\
PN G052.0+01.7 & ... & E & ... & ... & ... & L & MIRS\\
PN G051.7+01.3 & OAN-SPM, 2010, 2.1m, 1$\times$1800s & Ia$\ddagger$  & 4322743914667694208 & ... & ... & T & PN\\
PN G057.9+01.7 & ... & R & 2019455895594998784 & $5556_{-2409}^{+10,502}$ & $6174_{-2826}^{+2448}$ & T & PNC\\
PN G058.6+00.9 & ... & S & 2020219403339189760 & ... & ... & P & PN\\
PN G073.6+02.8 & KPNO, 2009, 2.1m, 1$\times$1800s & Eas$\dagger$  & ... & ... & ... & T & PN\\
PN G077.4$-$04.0 & ... & Es$\ddagger$  & ... & ... & ... & T & PN\\
PN G086.2$-$01.2 & ... & B  & ... & ... & ... & T & PNC\\
PN G095.8+02.6 & GTC, 2011, 10.4m, 1$\times$2400s & Eps$\ddagger$  & ... & ... & ... & T & PN\\
PN G098.9$-$01.1 & OAN-SPM, 2011, 2.1m, 1$\times$1800s & Ra$\ddagger$ & 1981477771828655488 & $5583_{-2607}^{+2964}$ & $5501_{-808}^{+1486}$ & T & PN\\
PN G114.4+00.0 & ... & E$\ddagger$ & 2012536875076571776 & $3367_{-1197}^{+2219}$ & $4251_{-920}^{+1219}$ & T & PN\\
\hline
\multicolumn{8}{l}{\small$^a$Observatory, date, telescope diameter, and exposure time. Spectra that are not flux-calibrated were not included.}\\
\multicolumn{8}{l}{\small$^b$Morphology. Round (R), elliptical (E), bipolar (B), star$-$like (S), irregular (I), sided asymmetry (a), multiple shells (m), internal structure (s), and}\\
\multicolumn{8}{l}{\small point symmetry (p). $\ast$\citet{2006MNRAS.373...79P}, $\dagger$\citet{2014MNRAS.443.3388S}, and $\ddagger$\citet{2021MNRAS.508.1599S}. Morphologies without references assigned by HASH.}\\
\multicolumn{8}{l}{\small$^c$DR3 IDs (\citealp{2023AA...674A...1G}) according to \citet{2021AA...656A.110C}, and \citet{2021AA...656A..51G}.}\\
\multicolumn{8}{l}{\small$^d$Geometric distance ($d_\mathrm{geo}$) and photogeometric distance ($d_\mathrm{phot}$) calculated by \citet{2021AJ....161..147B}.}\\
\multicolumn{8}{l}{\small$^e$HASH status. For Planetary Nebula (PN): Possible (P), True (T), and Likely (L). Other objects: Not PN (NPN) and Possible galaxy (PG).}\\
\multicolumn{8}{l}{\small$^f$SIMBAD status. PN Candidate (PNC), Star (S), Emission-line Star (ELS), and MIRS (Mid-IR Source).}
\end{tabular}   
\end{center}
\end{changemargin}
\end{table}  
\end{landscape} 

\newpage

\begin{landscape}
\begin{table}  
\begin{changemargin}{-1cm}{-1cm}
\begin{center} 
\vskip -1.5cm
\caption{Observed objects} \label{tab2}
\begin{tabular}{lccccc}
\hline 
Name & RA (2000) & DEC (2000) & $\mathrm{t_{exp} \, (s)}$ & OBSERVAT & DATE\\ 
\hline 
PN G003.3+66.1 & 14:16:21.95 & +13:52:24.25 & 3$\times$1800 & OAN-SPM & 2022 May 24\\
PN G339.4+29.7 & 15:08:20.73 & $-$23:14:50.47 & 1$\times$1800 &  OAN-SPM & 2022 May 24\\
PN G358.5+09.1 & 17:07:45.52 & $-$25:03:38.87 & 1$\times$1800&  OAN-SPM & 2022 July 27\\
PN G358.4+08.9 & 17:08:19.73 & $-$25:13:41.60 & 1$\times$1800&  OAN-SPM & 2022 July 26\\
PN G006.5+08.7$^a$ & 17:28:14.05 & $-$18:44:31.01 & 3$\times$1800 &  OAN-SPM & 2022 May 24\\
PN G008.3+09.6 & 17:29:13.10 & $-$16:47:42.60 & 3$\times$1800  &  OAN-SPM & 2022 May 24\\
PN G031.3+02.0 & 18:41:25.95 & $-$00:27:45.41 & 1$\times$1200  &  OAGH & 2016 May 07\\
PN G030.5+01.5 & 18:41:40.43 & $-$01:25:17.72 & 1$\times$1800  &  OAGH & 2016 May 07\\
PN G040.1+03.2 & 18:53:09.40 & +07:52:41.00 & 3$\times$1800  &  OAGH & 2016 May 08\\
PN G038.4+02.2 & 18:53:21.74 & +05:56:42.12 & 1$\times$1800  &  OAGH & 2016 May 08\\
PN G040.6+01.5 & 18:59:57.00 & +07:35:44.00 & 1$\times$1800  &  OAGH & 2016 May 10\\
PN G041.5+01.7 & 19:01:05.72 & +08:25:35.93 & 1$\times$1200  &  OAGH & 2016 May 08\\
PN G047.8+02.4 & 19:10:01.10 & +14:22:02.00 & 1$\times$1800  &  OAN-SPM & 2022 July 28\\
PN G050.0+01.0 &19:19:22.92 & +15:41:37.15 & 1$\times$1800  &  OAGH & 2016 May 09\\
PN G051.8+01.7 & 19:20:31.62 & +17:32:48.92 & 1$\times$1800  &  OAGH & 2016 May 09\\
PN G052.0+01.7 & 19:20:54.53 & +17:46:07.90 & 1$\times$1800 &  OAGH & 2016 May 09\\
PN G051.7+01.3 & 19:21:46.61 & +17:20:45.80 & 1$\times$1800 &  OAN-SPM & 2022 July 26\\
PN G057.9+01.7 & 19:33:09.03 & +22:58:33.63 & 1$\times$1800 &  OAGH & 2016 May 10\\
PN G058.6+00.9 & 19:37:29.33 & +23:09:46.56 & 1$\times$1800  &  OAGH & 2016 May 10\\
PN G073.6+02.8 & 20:05:22.00 & +36:59:42.00 & 1$\times$1800 &  OAN-SPM & 2022 July 28\\
PN G077.4$-$04.0 & 20:44:14.10 & +36:07:37.00 & 1$\times$1800  &  OAN-SPM & 2022 July 28\\
PN G086.2$-$01.2 & 21:02:38.73 & +44:46:46.00 & 1$\times$1800 &  OAN-SPM & 2022 July 28\\
PN G095.8+02.6 & 21:26:08.30 & +54:20:15.00 & 1$\times$1800  &  OAN-SPM & 2022 July 26\\
PN G098.9$-$01.1 & 21:58:42.30 & +53:30:03.00 & 1$\times$1800  &  OAN-SPM & 2022 July 26\\
PN G114.4+00.0 & 23:38:40.43 & +61:41:40.90 & 1$\times$1800 &  OAN-SPM & 2022 July 26\\ \hline
\multicolumn{6}{l}{\small$^a$Only object observed with $\mathrm{PA=144\arcdeg}$.}
\end{tabular} 
\end{center}
\end{changemargin}
\end{table}  
\end{landscape} 

\newpage  

\begin{landscape}
\begin{table} 
\begin{changemargin}{-1cm}{-1cm}
\begin{center}  
\vskip -2.5cm
\caption{Intrinsic line intensities $\mathrm{(I(H{\beta})=100)}$ and physical parameters for the first set of objects}
\label{tab3}
\vskip 0.1cm
\begin{tabular}{lcccccccc} \hline                              
Ion & Line & $f_{\lambda}$ & 003.3+66.1 & 006.5+08.7 & 008.3+09.6 & 030.5+01.5 & 047.8+02.4 & 057.9+01.7 \\
\hline
$\mathrm{H{\delta}}$ & 4101 & 0.230 & 8 $\pm$ 1 & ... & 27 $\pm$ 1 & ...  & ... & ... \\
$\mathrm{H{\gamma}}$ & 4340 & 0.157 & 48 $\pm$ 1 & 52 $\pm$ 1 & 48 $\pm$ 1 & ...  & 51 $\pm$ 2 & ... \\
$\ion{He}{I}$ & 4471 & 0.115 & ... & 6 $\pm$ 1 & 6 $\pm$ 1 & ...  & ... & ... \\
$\ion{He}{II}$ & 4686 & 0.050 & 55 $\pm$ 1 & 30 $\pm$ 1 & ... & ... & 6 $\pm$ 1 & ... \\
$\mathrm{H{\beta}}$ & 4861 & 0.000 & 100 $\pm$ 1 & 100 $\pm$ 1 & 100 $\pm$ 1 & 100 $\pm$ 1 & 100 $\pm$ 2 & 100 $\pm$ 1\\
$[\ion{O}{III}]$ & 4959 & $-$0.026 & 141 $\pm$ 2 & 236 $\pm$ 2 & 58 $\pm$ 1 & 409 $\pm$ 2 & 243 $\pm$ 4 & 165 $\pm$ 1\\
$[\ion{O}{III}]$ & 5007 & $-$0.038 & 366 $\pm$ 3 & 662 $\pm$ 6 & 172 $\pm$ 1 & 1169 $\pm$ 4 & 707 $\pm$ 10 & 471 $\pm$ 4\\
$[\ion{N}{II}]$ & 5755 & $-$0.185 & ... & ... & ... & ... &  6 $\pm$ 1 & ... \\
$\ion{He}{I}$ & 5876 & $-$0.203 & ... & 12 $\pm$ 1 & 15 $\pm$ 1 & 20 $\pm$ 1 & 14 $\pm$ 1 & 17 $\pm$ 1\\
$[\ion{O}{I}]$ & 6300 & $-$0.263 & ... & ... & ... & ... &  48 $\pm$ 1 & ... \\
$[\ion{O}{I}]$ & 6363 & $-$0.271 & ... & ... & ... & ... &  17 $\pm$ 1 & ... \\
$[\ion{N}{II}]$ & 6548 & $-$0.296 & ... & 48 $\pm$ 1 & 28 $\pm$ 1 & ... &  143 $\pm$ 3 & ... \\
$\mathrm{H{\alpha}}$ & 6563 & $-$0.298 & 286 $\pm$ 4 & 287 $\pm$ 4 & 286 $\pm$ 2 & 286 $\pm$ 2 & 286 $\pm$ 6 & 286 $\pm$ 3 \\
$[\ion{N}{II}]$ & 6583 & $-$0.300 & ... & 130 $\pm$ 2 & 79 $\pm$ 1 & 5 $\pm$ 1 & 443 $\pm$ 9 & 47 $\pm$ 1 \\
$\ion{He}{I}$ & 6678 & $-$0.313 & ... & 5 $\pm$ 1 & 4 $\pm$ 1 & 6 $\pm$ 1 &  5 $\pm$ 1 & 5 $\pm$ 1 \\
$[\ion{S}{II}]$ & 6716 & $-$0.318 & ... & 25 $\pm$ 1 & 17 $\pm$ 1 & ... &  51 $\pm$ 1 & 9 $\pm$ 1$^a$ \\
$[\ion{S}{II}]$ & 6731 & $-$0.320 & ... & 15 $\pm$ 1 & 13 $\pm$ 1 & ... &  46 $\pm$ 1 & ... \\
$\ion{He}{I}$ & 7065 & $-$0.364 & ... & ... & ... & ... &  5 $\pm$ 1 & 7 $\pm$ 1 \\
$[\ion{Ar}{III}]$ & 7136 & $-$0.374 & ... & 15 $\pm$ 1 & 7 $\pm$ 1 & ... & 22 $\pm$ 1 & 19 $\pm$ 1 \\
$[\ion{O}{II}]$ & 7320 & $-$0.398 & ... & ... & ... & ... & 5 $\pm$ 1 & 5 $\pm$ 1$^b$ \\
$[\ion{O}{II}]$ & 7330 & $-$0.400 & ... & ... & ... & ... &  12 $\pm$ 1 & ... \\
\hline \hline
$\log \, F(\mathrm{H\beta})$ & --- & --- & $-$14.19 $\pm$ 0.01 & $-$14.26 $\pm$ 0.01 & $-$13.70 $\pm$ 0.01 & $-$14.22 $\pm$ 0.01 & $-$14.50 $\pm$ 0.01 & $-$14.58 $\pm$ 0.01 \\
$SB(\mathrm{H\beta})^c$ & --- & --- & 2.75 & 4.62 & 19.40 & 26.37 & 9.05 & 15.73 \\
$c(\mathrm{H\beta})$ & --- & --- & 0.09 $\pm$ 0.01 & 1.26 $\pm$ 0.01 & 0.96 $\pm$ 0.01 & 3.33 $\pm$ 0.01 & 1.83 $\pm$ 0.02 & 3.29 $\pm$ 0.01 \\
$T_\mathrm{e}([\ion{N}{II}])$& --- & --- & .... & ... & ... & ... & 10,295 $\pm$ 258 & ... \\
$n_\mathrm{e}([\ion{S}{II}])$& --- & --- & ... & $<100$ & 147 $\pm$ 33 & ... & 381 $\pm$ 90 & ... \\
\hline 
\multicolumn{9}{l}{\small$^a$Correspond to $\lambda$6716 and $\lambda$6731 sulfur lines blended.}\\
\multicolumn{9}{l}{\small$^b$Correspond to $\lambda$7320 and $\lambda$7330 oxygen lines blended.}\\
\multicolumn{9}{l}{\small$^c$Surface brightness $(\times10^{-17}\mathrm{erg\,s^{-1}\,cm^{-2}\,arcsec^{-2}})$.}
\end{tabular}  
\end{center}
\end{changemargin}
\end{table}  
\end{landscape} 

\newpage

\begin{landscape}
\begin{table}  
\begin{changemargin}{-1cm}{-1cm}
\begin{center} 
\vskip -1.7cm
\caption{Intrinsic line intensities $\mathrm{(I(H{\beta})=100)}$ and physical parameters for the second set of objects}
\label{tab4}
\vskip 0.1cm
\begin{tabular}{lccccccc} \hline                         
Ion & Line & $f_{\lambda}$ & 073.6+02.8 & 077.4$-$04.0 & 086.2$-$01.2 & 098.9$-$01.1 & 114.4+00.0 \\
\hline
$\mathrm{H{\delta}}$ & 4101 & 0.230 & 84 $\pm$ 6 & ... & 34 $\pm$ 1  & ... & ... \\
$\mathrm{H{\gamma}}$ & 4340 & 0.157  & ... & ... & 66 $\pm$ 1 & ... & ... \\
$\ion{He}{I}$ & 4471 & 0.115  & ... & ... & ... & ... . & ... \\
$\ion{He}{II}$ & 4686 & 0.050  & ... & ... & 19 $\pm$ 1 &  ... & ... \\
$\mathrm{H{\beta}}$ & 4861 & 0.000 & 100 $\pm$ 5 & 100 $\pm$ 4 & 100 $\pm$ 1 & 100 $\pm$ 4 & 100 $\pm$ 3\\
$[\ion{O}{III}]$ & 4959 & $-$0.026 & 230 $\pm$ 9 & 66 $\pm$ 3 & 278 $\pm$ 3 & 276 $\pm$ 9 & 193 $\pm$ 4 \\
$[\ion{O}{III}]$ & 5007 & $-$0.038 & 663 $\pm$ 23 & 216 $\pm$ 8 & 830 $\pm$ 8 & 851 $\pm$ 26 & 535 $\pm$ 11 \\
$[\ion{N}{II}]$ & 5755 & $-$0.185 & ... & ... & 6 $\pm$ 1  & ... & ... \\
$\ion{He}{I}$ & 5876 & $-$0.203 & 22 $\pm$ 1 & 23 $\pm$ 1 & 17 $\pm$ 1 & ... & ... \\
$[\ion{O}{I}]$ & 6300 & $-$0.263 & 42 $\pm$ 2 & 34 $\pm$ 2 & 22 $\pm$ 1 & 65 $\pm$ 2 & 88 $\pm$ 3 \\
$[\ion{O}{I}]$ & 6363 & $-$0.271 & 19 $\pm$ 1 & 7 $\pm$ 1 & 8 $\pm$ 1 & ... & ... \\
$[\ion{N}{II}]$ & 6548 & $-$0.296 & 110 $\pm$ 5 & 241 $\pm$ 11 & 109 $\pm$ 2 & 60 $\pm$ 3 & 66 $\pm$ 2 \\
$\mathrm{H{\alpha}}$ & 6563 & $-$0.298 & 286 $\pm$ 14 & 286 $\pm$ 13 & 285 $\pm$ 4 & 287 $\pm$ 13 & 286 $\pm$ 8 \\
$[\ion{N}{II}]$ & 6583 & $-$0.300 & 337 $\pm$ 16 & 714 $\pm$ 32 & 336 $\pm$ 5 & 161 $\pm$ 7 & 198 $\pm$ 6 \\
$\ion{He}{I}$ & 6678 & $-$0.313 & ... & ... & 5 $\pm$ 1 & ... & ... \\
$[\ion{S}{II}]$ & 6716 & $-$0.318 & 57 $\pm$ 3 & 43 $\pm$ 2 & 27 $\pm$ 1 & 43 $\pm$ 2 & 33 $\pm$ 1 \\
$[\ion{S}{II}]$ & 6731 & $-$0.320 & 47 $\pm$ 3 & 34 $\pm$ 2 & 19 $\pm$ 1 & 16 $\pm$ 1 & 26 $\pm$ 1 \\
$\ion{He}{I}$ & 7065 & $-$0.364 & ... & ... & 4 $\pm$ 1 & ... & ... \\
$[\ion{Ar}{III}]$ & 7136 & $-$0.374 & 16 $\pm$ 1 & ... & 19 $\pm$ 1 & ... & ... \\
$[\ion{O}{II}]$ & 7320 & $-$0.398 & ... & ... & 4 $\pm$ 1 & ... & ... \\
$[\ion{O}{II}]$ & 7330 & $-$0.400 & ... & ... & 4 $\pm$ 1 & ... & ... \\
\hline \hline
$\log \, F(\mathrm{H\beta})$ & --- & --- & $-$15.29 $\pm$ 0.01 & $-$15.24 $\pm$ 0.01 & $-$14.18 $\pm$ 0.01 & $-$15.20 $\pm$ 0.01 & $-$14.83 $\pm$ 0.01 \\
$SB(\mathrm{H\beta})^a$ & --- & --- & 1.44 & 0.92 & 10.87 & 0.03 & 1.11 \\
$c(\mathrm{H\beta})$ & --- & --- & 2.53 $\pm$ 0.05 & 1.78 $\pm$ 0.05 & 1.61 $\pm$ 0.01 & 1.75 $\pm$ 0.04 & 1.26 $\pm$ 0.03 \\
$T_\mathrm{e}([\ion{N}{II}])$ & --- & --- & ... & ... & 11,445 $\pm$ 216 & ... & ... \\
$n_\mathrm{e}([\ion{S}{II}])$ & --- & --- & 230 $\pm$ 170 & 175 $\pm$ 163& $<100$ & $<100$ & 155 $\pm$ 111 \\
\hline
\multicolumn{8}{l}{\small$^a$Surface brightness $(\times10^{-17}\mathrm{erg\,s^{-1}\,cm^{-2}\,arcsec^{-2}})$.}
\end{tabular}   
\end{center}
\end{changemargin}
\end{table}  
\end{landscape} 

\newpage

\begin{landscape}
\begin{table}  
\begin{changemargin}{-1cm}{-1cm}
\begin{center} 
\vskip -3.5cm
\caption{Mean ionic abundances\lowercase{$^a$} from all the selected objects} 
\label{tab5}
\vskip 0.1cm
\begin{tabular}{lccccccc} \hline    
Ion & Factor & 003.3+66.1 & 006.5+08.7 & 008.3+09.6 & 030.5+01.5 & 047.8+02.4 & 057.9+01.7\\
\hline 
He$^{1+}$ & $\times10^{-1}$ & ... & 1.09 $\pm$ 0.02 & 1.08 $\pm$ 0.01 & 1.55 $\pm$ 0.01 & 1.17 $\pm$ 0.02 & 1.28 $\pm$ 0.01\\
He$^{2+}$ & $\times10^{-2}$ & 4.48 $\pm$ 0.06 & 2.40 $\pm$ 0.04 & ... & ... & 0.47 $\pm$ 0.03 & ...\\
N$^{1+}$ & $\times10^{-5}$ & ... & 2.65 $\pm$ 0.03 & 1.56 $\pm$ 0.01 & 0.09 $\pm$ 0.01 & 7.94 $\pm$ 0.47 & 0.92 $\pm$ 0.01\\
O$^{0+}$ & $\times10^{-4}$ & ... & ... & ... & ... & 0.86 $\pm$ 0.07 & ...\\
O$^{1+}$ & $\times10^{-4}$ & ... & ... &  ... & ... & 2.57 $\pm$ 0.50 & 1.95 $\pm$ 0.03\\
O$^{2+}$ & $\times10^{-4}$ & 1.36 $\pm$ 0.01 & 2.39 $\pm$ 0.01 & 0.60 $\pm$ 0.01 & 4.18 $\pm$ 0.01 & 2.26 $\pm$ 0.16 & 1.69 $\pm$ 0.01\\
S$^{1+}$ & $\times10^{-6}$ & ... & 0.89 $\pm$ 0.01 & 0.69 $\pm$ 0.01 & ... & 2.19 $\pm$ 0.17 & 0.34 $\pm$ 0.01\\
Ar$^{2+}$ & $\times10^{-6}$ & ... & 1.41 $\pm$ 0.03 & 0.62 $\pm$ 0.01 & ... & 1.91 $\pm$ 0.16 & 1.76 $\pm$ 0.02\\
\hline  
\end{tabular}  

\vspace{0.3cm}

\begin{tabular}{lcccccc} \hline    
Ion & Factor & 073.6+02.8 & 077.4$-$04.0 & 086.2$-$01.2 & 098.9$-$01.1 & 114.4+00.0\\
\hline  
He$^{1+}$ & $\times10^{-1}$  &1.58 $\pm$ 0.08 & 1.71 $\pm$ 0.09 & 1.32 $\pm$ 0.02 & ... & ... \\
He$^{2+}$ & $\times10^{-2}$ &... & ... & 1.56 $\pm$ 0.04 & ... & ... \\
N$^{1+}$ & $\times10^{-5}$ &6.52 $\pm$ 0.23 & 14.01 $\pm$ 0.49 & 4.65 $\pm$ 0.19 & 3.31 $\pm$ 0.11 & 3.86 $\pm$ 0.09 \\
O$^{0+}$ & $\times10^{-4}$ &0.94 $\pm$ 0.03 & 0.59 $\pm$ 0.02 & 0.27 $\pm$ 0.02 & 1.26 $\pm$ 0.05 & 1.69 $\pm$ 0.05 \\
O$^{1+}$ & $\times10^{-4}$ &... & ... & 0.81 $\pm$ 0.08 & ... & ... \\
O$^{2+}$ & $\times10^{-4}$ & 2.35 $\pm$ 0.07 & 0.74 $\pm$ 0.02 & 1.88 $\pm$ 0.09 & 2.97 $\pm$ 0.07 & 1.93 $\pm$ 0.03 \\
S$^{1+}$ & $\times10^{-6}$ & 2.47 $\pm$ 0.19 & 1.79 $\pm$ 0.14 & 0.75 $\pm$ 0.03 & 1.34 $\pm$ 0.05 & 1.37 $\pm$ 0.08 \\
Ar$^{2+}$ & $\times10^{-6}$ &1.47 $\pm$ 0.09 & ... & 1.29 $\pm$ 0.08 & ... & ... \\
\hline  
\multicolumn{7}{l}{\small$^a$Given as $N(\mathrm{X^{i+}})/N(\mathrm{H^{+}})$.}
\end{tabular} 
\vspace{0.5cm}

\caption{Elemental abundances\lowercase{$^a$} from some selected objects} 
\label{tab6}
\begin{tabular}{lcccccccc} \hline    
Ion & Factor & 006.5+08.7 & 047.8+02.4 & 086.2$-$01.2 & Sun$^b$ & KB94$^c$& SGC06$^c$ & TPZ24$^c$ \\
\hline  
He & 1 & 0.133 $\pm$ 0.002 & 0.122 $\pm$ 0.003 & 0.147 $\pm$ 0.002 & 0.082 & 0.115 (55)& 0.123 (75) & 0.115 (123)\\
N & $\times10^{-4}$ &...&1.532 $\pm$ 0.357&1.660 $\pm$ 0.193 & 0.676& 2.239 (47)& 2.439 (63) & 1.862 (122)\\
O & $\times10^{-4}$ &...& 4.964 $\pm$ 0.548 & 2.904 $\pm$ 0.134 & 4.898& 4.786 (54)& 3.531 (67) & 5.012 (124)\\
S & $\times10^{-5}$ &...&1.511 $\pm$ 0.124&0.626 $\pm$ 0.041 & 1.318& 0.832 (43)& \,\, ...  \, (...) & 0.794 (124)\\
Ar & $\times10^{-6}$ & 2.648 $\pm$ 0.583 & 3.574 $\pm$ 0.840 &2.420 $\pm$ 0.550 & 2.399& 2.455 (41)& 1.262 (46) & 2.754 (122)\\
\hline   
\multicolumn{9}{l}{\small$^a$Given as $N(\mathrm{X})/N(\mathrm{H})$.}\\
\multicolumn{9}{l}{\small$^b$According to \citet{2021AA...653A.141A}.}\\
\multicolumn{9}{l}{\small$^c$Average PNe abundances. KB94 is for \citet{1994MNRAS.271..257K}, SGC06 for \citet{2006ApJ...651..898S}, and}\\
\multicolumn{9}{l}{\small TPZ24 for \citet{2024MNRAS.527.6363T}. The number of PNe used for each average appear in parentheses.}
\end{tabular} 
\end{center}
\end{changemargin}
\end{table}  
\end{landscape} 

\newpage

\begin{landscape}
\begin{table}  
\begin{changemargin}{-1cm}{-1cm}
\begin{center} 
\vskip -2.5cm
\caption{Objects with few emission lines detected} 
\label{tab7}
\begin{tabular}{lccccccc} \hline    
Ion & Line & 031.3+02.0 & 040.1+03.2$^a$ & 038.4+02.2 & 040.6+01.5 & 041.5+01.7 & 050.0+01.0 \\
\hline  
$[\ion{O}{III}]$ & 4959 &  ... &  13 $\pm$ 1 & ... & ... & ... & ...\\
$[\ion{O}{III}]$ & 5007 & ... & 30 $\pm$ 1 & 19 $\pm$ 1 & ... & ... & 24 $\pm$ 1 \\
$[\ion{N}{II}]$ & 6548 &  ... &  ... & ... & ... & ... & ...\\
$\mathrm{H{\alpha}}$ & 6563 & 100 $\pm$ 1 &  100 $\pm$ 1 & 100 $\pm$ 1 & 100 $\pm$ 1 & 100 $\pm$ 1$^b$ & 100 $\pm$ 1\\
$[\ion{N}{II}]$ & 6583 &  ... &  4 $\pm$ 1 & ... & ... & 40 $\pm$ 1 & ...\\
$[\ion{S}{II}]$ & 6716 &  ... &  ...& ... & ... & ... & ...\\
$[\ion{S}{II}]$ & 6731 &  ... &  ...& ... & ... & ... & ...\\
$[\ion{Ar}{III}]$ & 7136 &  ... &  ...& ... & ... & ... & ...\\
\hline \hline
$\log \, F(\mathrm{H\alpha})$ & --- & $-$14.29 $\pm$ 0.01 & $-$13.31 $\pm$ 0.01 & $-$13.80 $\pm$ 0.01 & $-$14.89 $\pm$ 0.01 & $-$14.19 $\pm$ 0.01 & $-$14.11 $\pm$ 0.01\\
$SB(\mathrm{H\alpha})^c$ & --- & 6.16 & 16.78 & 11.01 & 1.33 & 10.71 & 10.73 \\
$c(\mathrm{H\beta})_\mathrm{limit}$ & --- & $>0.81$ & $>3.04$ & $>1.82$ & ... & ... & $>1.67$ \\
\hline
\multicolumn{8}{l}{\small$^a$We found also $\mathrm{I(\ion{C}{IV}) = 16 \pm 1}$, see \S~\ref{sec:individual}.}\\
\multicolumn{8}{l}{\small$^b$Blended with $\mathrm{\lambda6548}$, see \S~\ref{sec:individual}.}\\
\multicolumn{8}{l}{\small$^c$Surface brightness $(\times10^{-16}\mathrm{erg\,s^{-1}\,cm^{-2}\,arcsec^{-2}})$.}

\end{tabular} 

\vspace{0.5cm}

\begin{tabular}{lcccccc} \hline    
Ion & Line & 051.8+01.7 & 052.0+01.7 & 051.7+01.3 & 058.6+00.9 & 095.8+02.6\\
\hline  
$[\ion{O}{III}]$ & 4959 &  ... &  ... & 15 $\pm$ 1 & ... & 12 $\pm$ 1\\
$[\ion{O}{III}]$ & 5007 & ... & ... &  43 $\pm$ 1 &  ... &  39 $\pm$ 1 \\
$[\ion{N}{II}]$ & 6548 &  ... &  ... & 29 $\pm$ 1 & ... & 33 $\pm$ 1\\
$\mathrm{H{\alpha}}$ & 6563 & 100 $\pm$ 1 & 100 $\pm$ 1 & 100 $\pm$ 2 & 100 $\pm$ 3 &  100 $\pm$ 1\\
$[\ion{N}{II}]$ & 6583 & ... & ... & 97 $\pm$ 2 & ...&  110 $\pm$ 2\\ 
$[\ion{S}{II}]$ & 6716 &  ... &  ... & 16 $\pm$ 1 & ... & 12 $\pm$ 1 \\
$[\ion{S}{II}]$ & 6731 &  ... &  ... & 12 $\pm$ 1 & ... & 8 $\pm$ 1 \\
$[\ion{Ar}{III}]$ & 7136 &  ... &  ... & 23 $\pm$ 1 & ... & 14 $\pm$ 1 \\
\hline \hline
$\log \, F(\mathrm{H\alpha})$ & --- & $-$14.82 $\pm$ 0.01 & $-$13.72 $\pm$ 0.01 & $-$14.42 $\pm$ 0.01 & $-$15.40 $\pm$ 0.01& $-$14.19 $\pm$ 0.01\\
$SB(\mathrm{H\alpha})^a$ &  ---  & 2.11 & 13.12 & 0.53 & 0.66 & 1.60 \\ 
$c(\mathrm{H\beta})_\mathrm{limit}$ & --- & ... & $>1.08$ & $>2.31$ & ... & $>3.16$ \\
\hline
\multicolumn{7}{l}{\small$^a$Surface brightness $(\times10^{-16}\mathrm{erg\,s^{-1}\,cm^{-2}\,arcsec^{-2}})$.}
\end{tabular}
\end{center}
\end{changemargin}
\end{table}  
\end{landscape} 

\newpage

\begin{landscape}
\begin{table}  
\begin{changemargin}{-1cm}{-1cm}
\begin{center} 
\vskip -2.5cm
\caption{Object sizes\lowercase{$^a$} from our sample} \label{tab8}
\begin{tabular}{lccc}
\hline 
Name & [\ion{O}{III}] 5007 & $\mathrm{H{\alpha}}$ 6563 & Literature major diameter \\
 & (arcsec) & (arcsec) & (arcsec) \\
\hline
PN G003.3+66.1 & 106 & 100 & 50 \citep{2016MNRAS.455.1459F} \\
PN G339.4+29.7 & ... & ... & ...\\
PN G358.5+09.1 & ... & ... & ... \\
PN G358.4+08.9 & ... & ... & ... \\
PN G006.5+08.7 & 47 & 45 & 51 \citep{2006MNRAS.373...79P}\\
PN G008.3+09.6 & 29 & 38 & 30 \citep{2006MNRAS.373...79P}\\
PN G031.3+02.0 & ... & 3 & ...\\
PN G030.5+01.5 & 8 & 11 & ...\\
PN G040.1+03.2 & 5 & 7 & 12 \citep{2014MNRAS.443.3388S}\\
PN G038.4+02.2 & 4 & 4 & 5 \citet{2014MNRAS.443.3388S}\\
PN G040.6+01.5 & ... & 3 & 5 \citep{2021MNRAS.508.1599S}\\
PN G041.5+01.7 & ... & 3 & ...\\
PN G047.8+02.4 & 10 & 12 & 10 \citep{2014MNRAS.443.3388S}\\
PN G050.0+01.0 & 3 & 3 & ...\\
PN G051.8+01.7 & ... & 3 & ...\\
PN G052.0+01.7 & ... & 5 & ...\\
PN G051.7+01.3 & 23 & 25 & 34 \citep{2021MNRAS.508.1599S}\\
PN G057.9+01.7 & 5 & 6 & ...\\
PN G058.6+00.9 & ... & 3 & ...\\
PN G073.6+02.8 & 10 & 12 & 11 \citep{2014MNRAS.443.3388S}\\
PN G077.4$-$04.0 & 14 & 23 & 25 \citep{2021MNRAS.508.1599S}\\
PN G086.2$-$01.2 & 15 & 17 & 12 \citet{2015LAstr.129a..42F}\\
PN G095.8+02.6 & 9 & 12 & 15 \citep{2021MNRAS.508.1599S}\\
PN G098.9$-$01.1 & 26 & 26 & 31 \citep{2021MNRAS.508.1599S}\\
PN G114.4+00.0 & 55 & 56 & 61 \citep{2021MNRAS.508.1599S}\\ \hline
\multicolumn{4}{l}{\small$^a$Uncertainties are determined by the seeing, with a typical value of 2 arcsec.}
\end{tabular} 
\end{center}
\end{changemargin}
\end{table}  
\end{landscape} 

\appendix

\begin{appendices}
\hypertarget{appendixA}{}
\section*{APPENDIX A: SLIT POSITION AND SPECTRUM EXTRACTION ZONE} 
\label{appendix:A}
\renewcommand{\thefigure}{A\arabic{figure}}
\setcounter{figure}{0}
\begin{figure}[H]
\begin{changemargin}{-4cm}{-4cm}
\begin{center}
\includegraphics[scale=0.6]{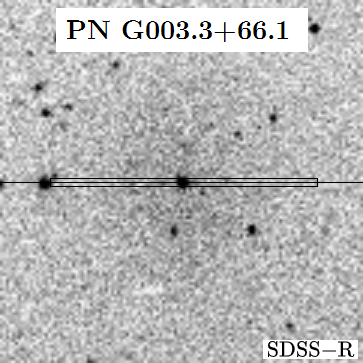} \includegraphics[scale=0.6]{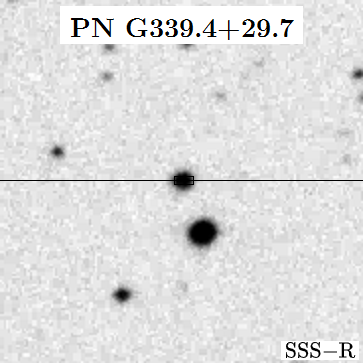} \includegraphics[scale=0.6]{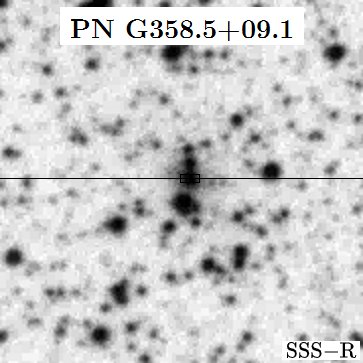}\\ 
\vspace{0.1 cm}
\includegraphics[scale=0.6]{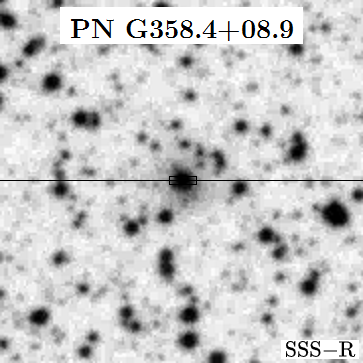} \includegraphics[scale=0.6]{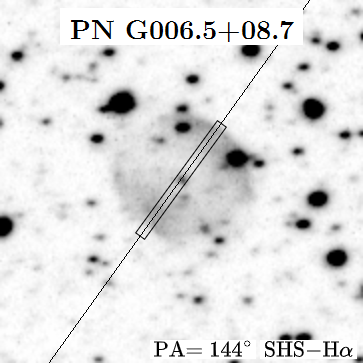} \includegraphics[scale=0.6]{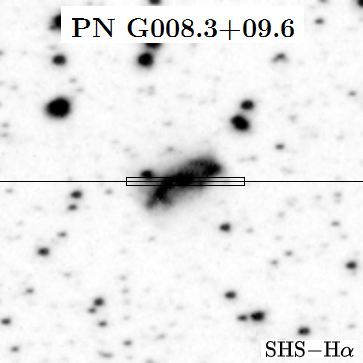}\\ 
\vspace{0.1 cm}
\includegraphics[scale=0.6]{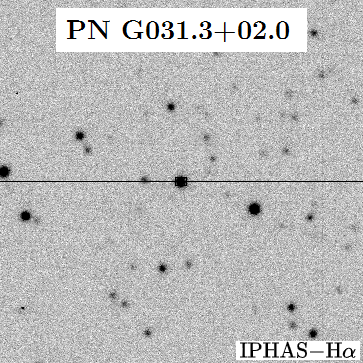} \includegraphics[scale=0.6]{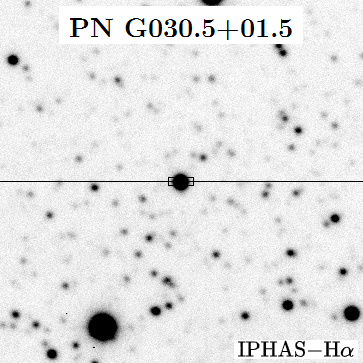} \includegraphics[scale=0.6]{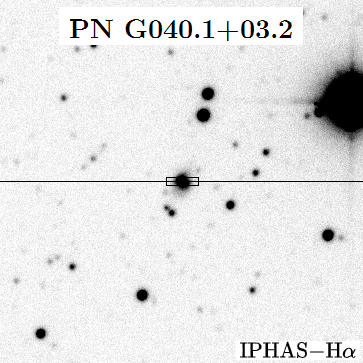}
\end{center}
\caption{Visual fields available in HASH for observed objects with images taken from SDSS \citep{2000AJ....120.1579Y}, SSS \citep{2001MNRAS.326.1279H}, SHS \citep{2005MNRAS.362..689P}, and IPHAS \citep{2005MNRAS.362..753D}, as labeled in the bottom right corner of each panel. North is up, East is left. Field size is 2{\arcmin}$\times$2{\arcmin} in all cases. Solid line represents the long-slit position which was oriented at $\mathrm{PA=90\arcdeg}$ for all objects except where indicated. The rectangle over the long-slit marks the extraction zone for the spectra shown in Figure \ref{fig:1} and Appendix \protect\hyperlink{appendixB}{B}.}
\end{changemargin}
\end{figure}

\newpage

\begin{figure}[H]
\addtocounter{figure}{-1}
\begin{changemargin}{-4cm}{-4cm}
\begin{center}
\includegraphics[scale=0.6]{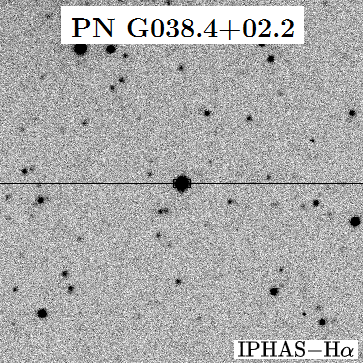} \includegraphics[scale=0.6]{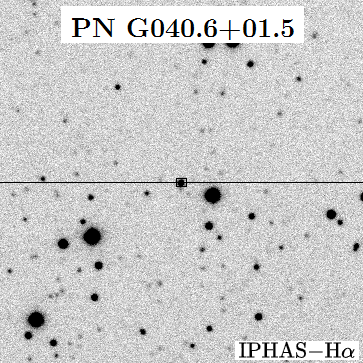} \includegraphics[scale=0.6]{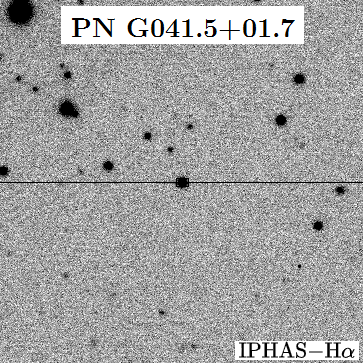}\\ 
\vspace{0.1 cm}
\includegraphics[scale=0.6]{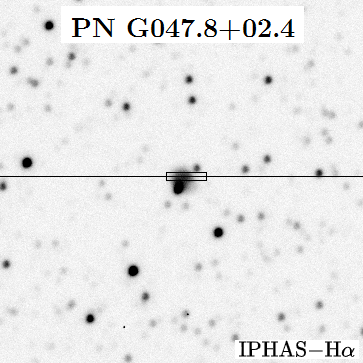} \includegraphics[scale=0.6]{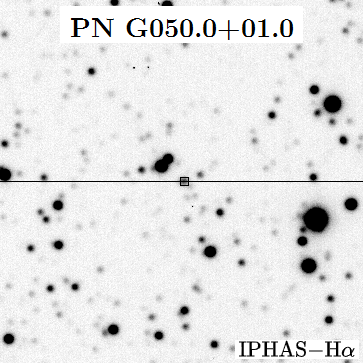} \includegraphics[scale=0.6]{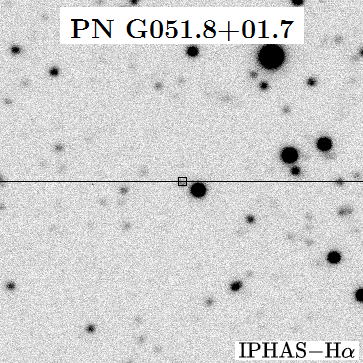}\\ 
\vspace{0.1 cm}
\includegraphics[scale=0.6]{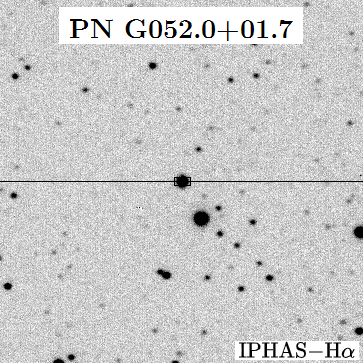} \includegraphics[scale=0.6]{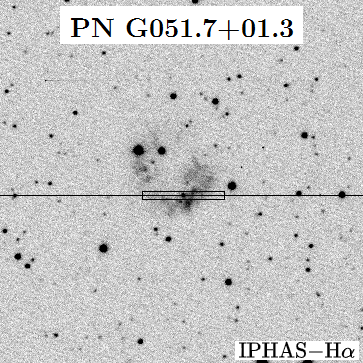} \includegraphics[scale=0.6]{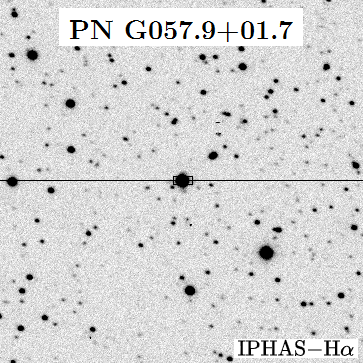}
\caption{(Continued).}
\end{center}
\end{changemargin}
\end{figure}

\newpage

\begin{figure}[H]
\addtocounter{figure}{-1}
\begin{changemargin}{-4cm}{-4cm}
\begin{center}
\includegraphics[scale=0.6]{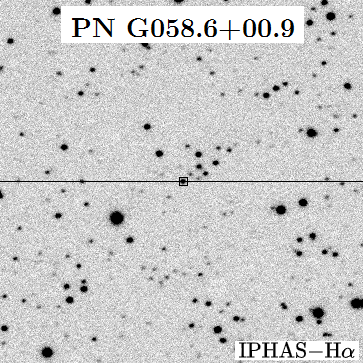} \includegraphics[scale=0.6]{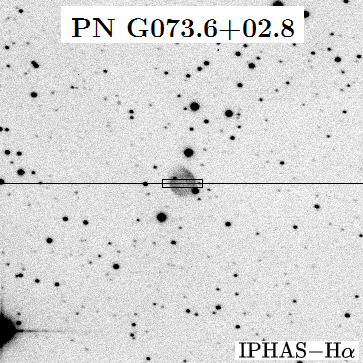} \includegraphics[scale=0.6]{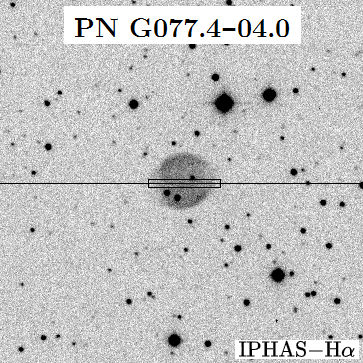}\\ 
\vspace{0.1 cm}
\includegraphics[scale=0.6]{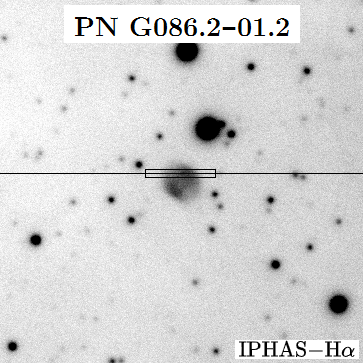} \includegraphics[scale=0.6]{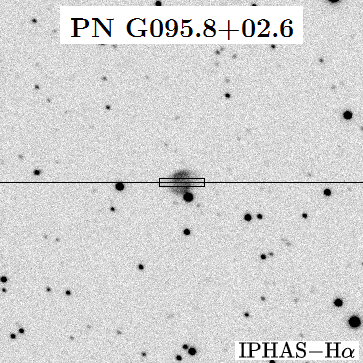} \includegraphics[scale=0.6]{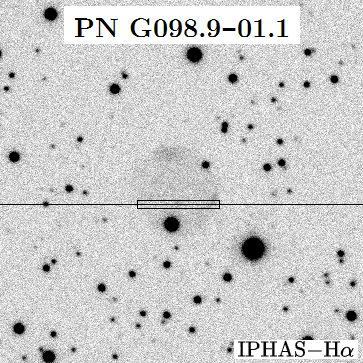}\\ 
\vspace{0.1 cm}
 \includegraphics[scale=0.6]{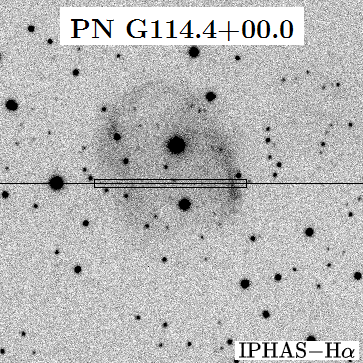}
\caption{(Continued).}
\end{center}
\end{changemargin}
\end{figure}

\hypertarget{appendixB}{}
\section*{APPENDIX B: SPECTRA OF THE REMAINING 14 OBJECTS}
\label{appendix:B}
\begin{figure}[H]
\begin{changemargin}{-3cm}{-3cm}
\begin{center}
\includegraphics[scale=0.35]{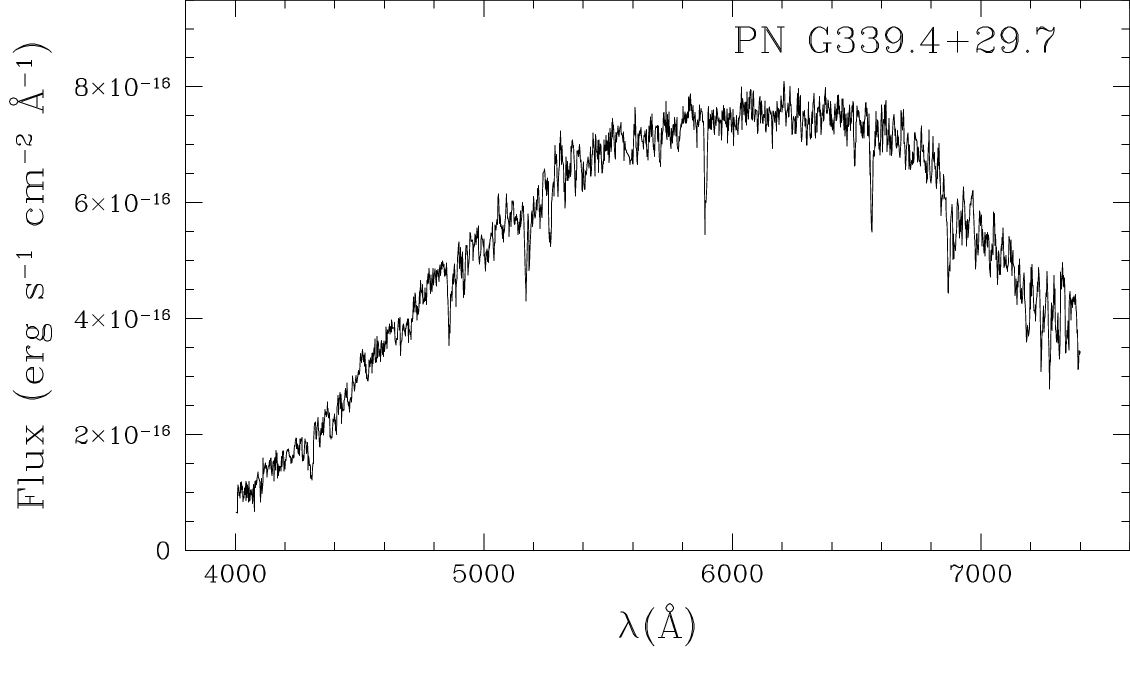} \hspace{0.3 cm} \includegraphics[scale=0.35]{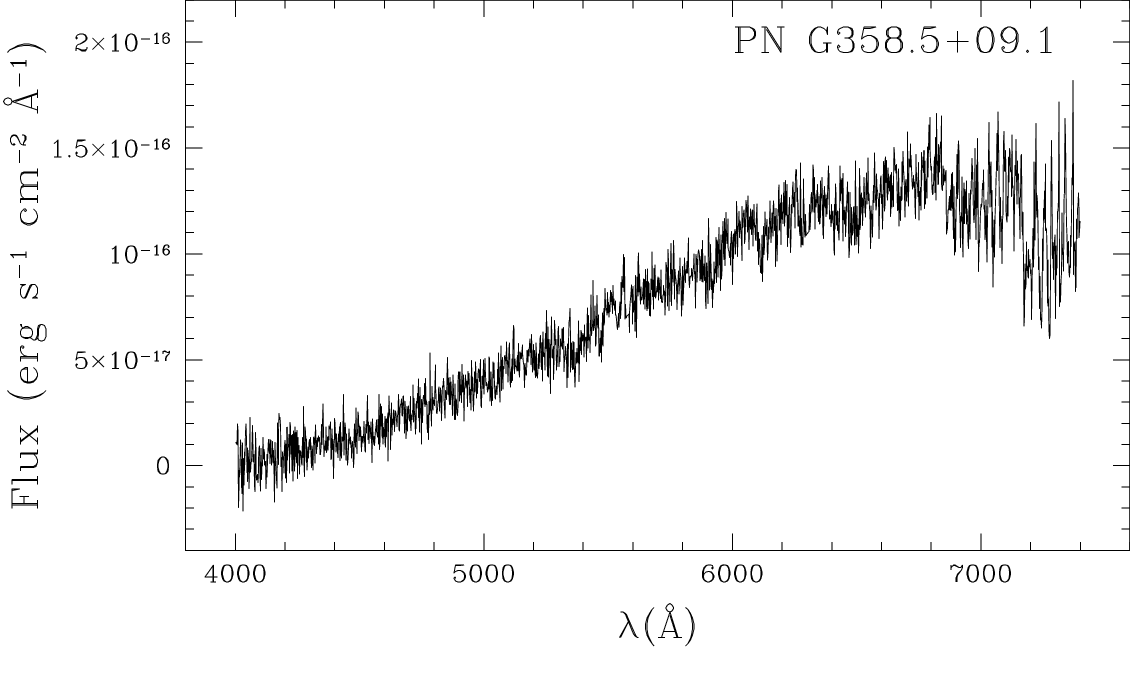}\\ \vspace{0.2 cm}
\includegraphics[scale=0.35]{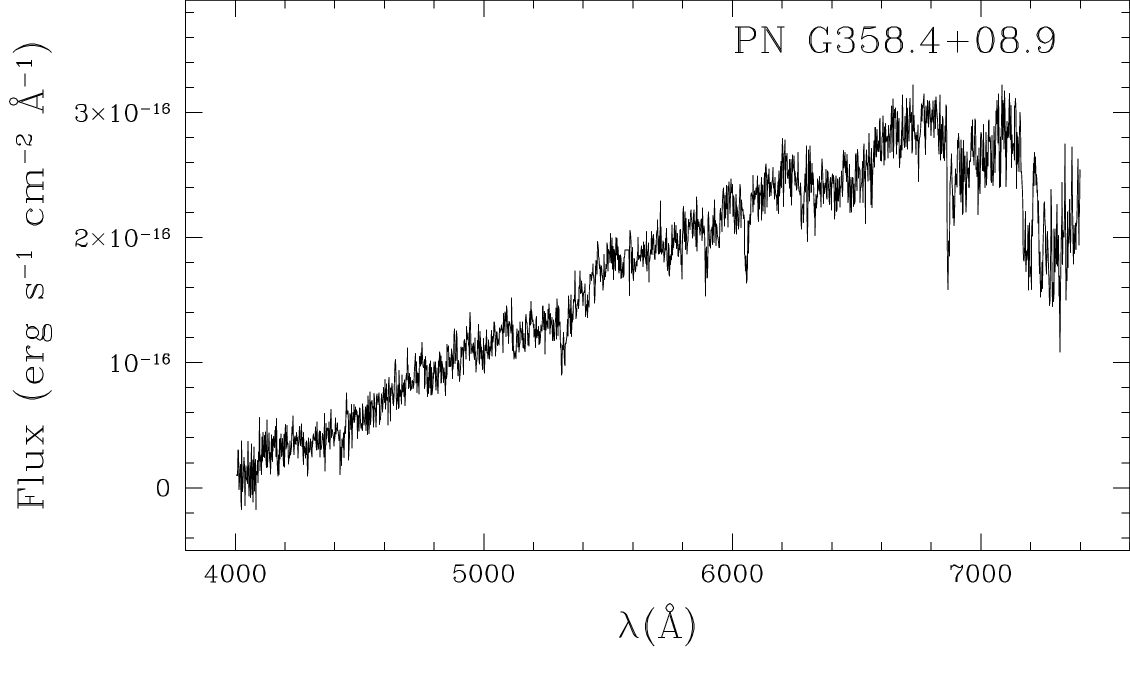} \hspace{0.3 cm} \includegraphics[scale=0.35]{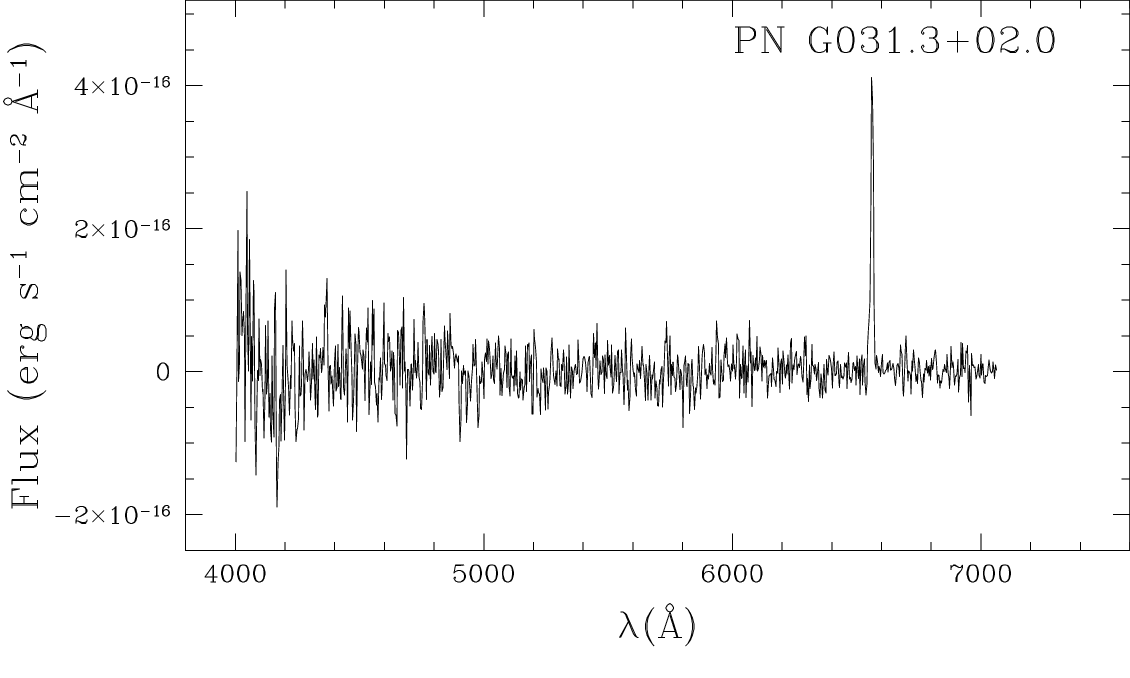}\\ \vspace{0.2 cm}
\includegraphics[scale=0.35]{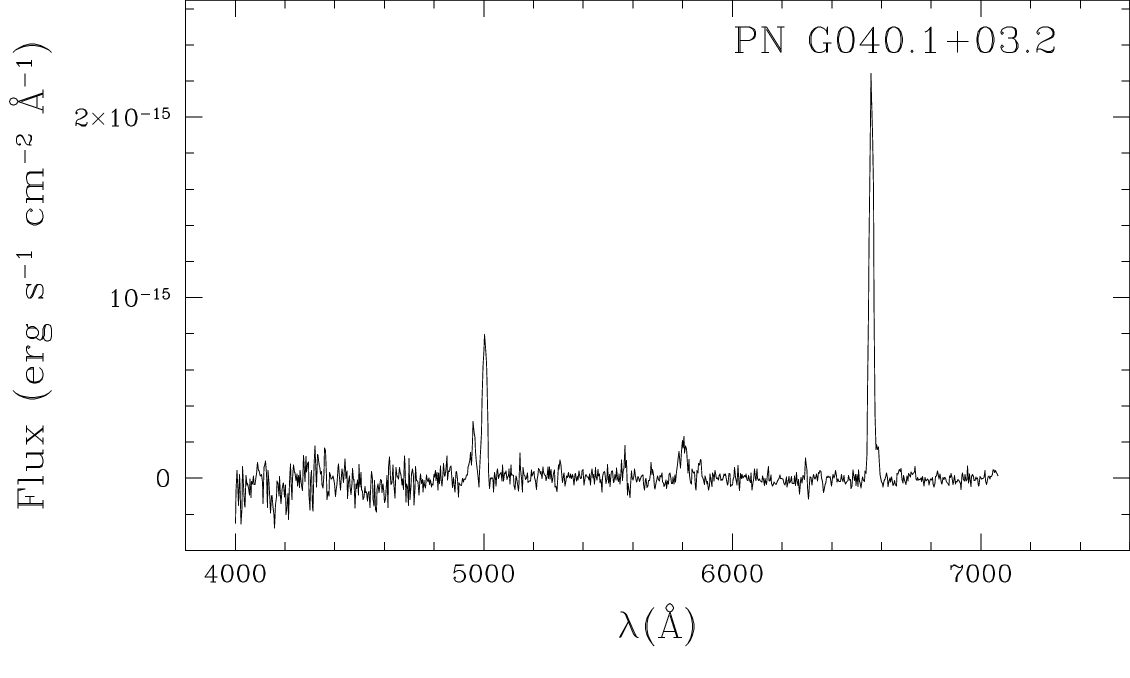} \hspace{0.3 cm} \includegraphics[scale=0.35]{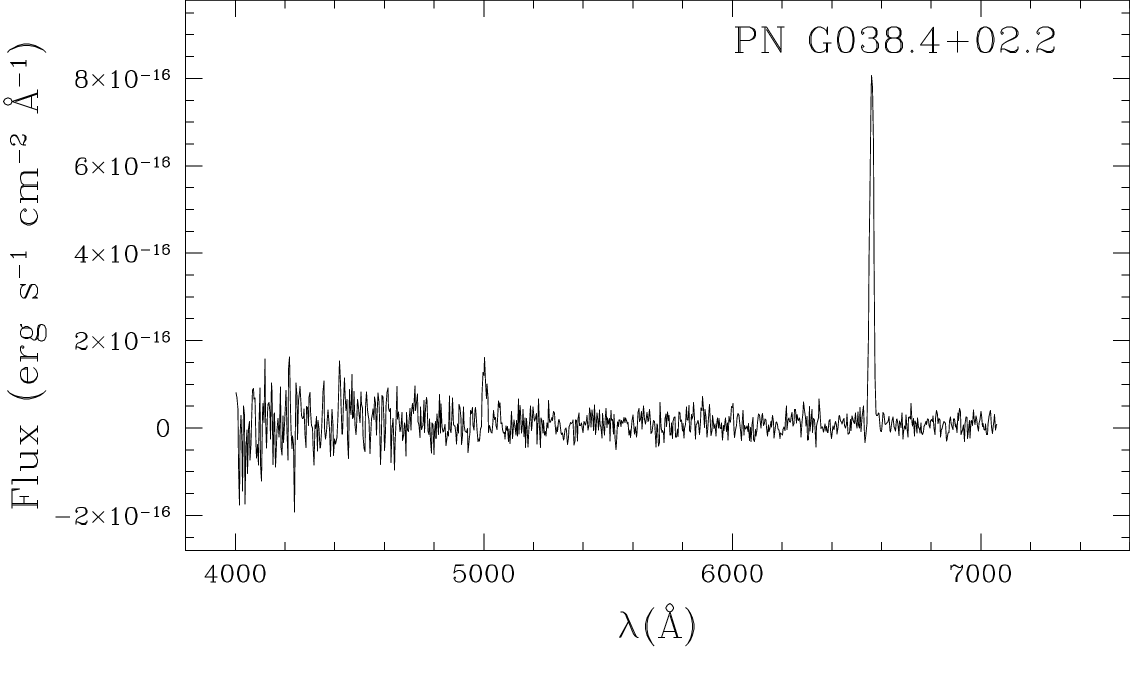}\\
\vspace{0.2 cm}
\includegraphics[scale=0.35]{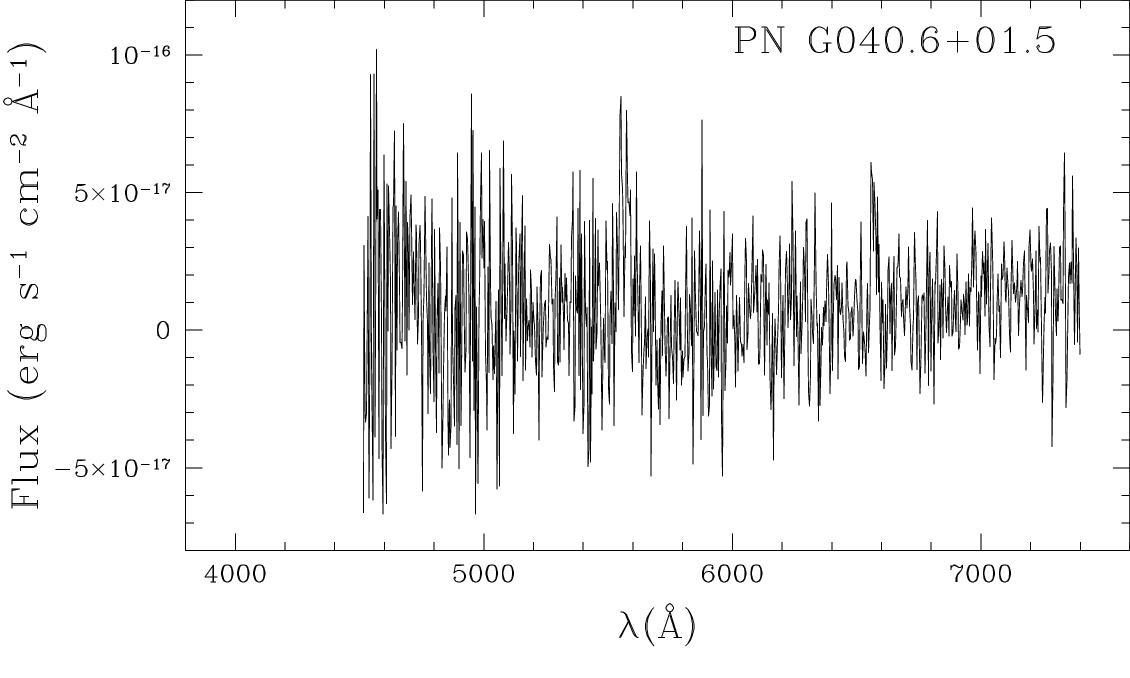} \hspace{0.3 cm} \includegraphics[scale=0.35]{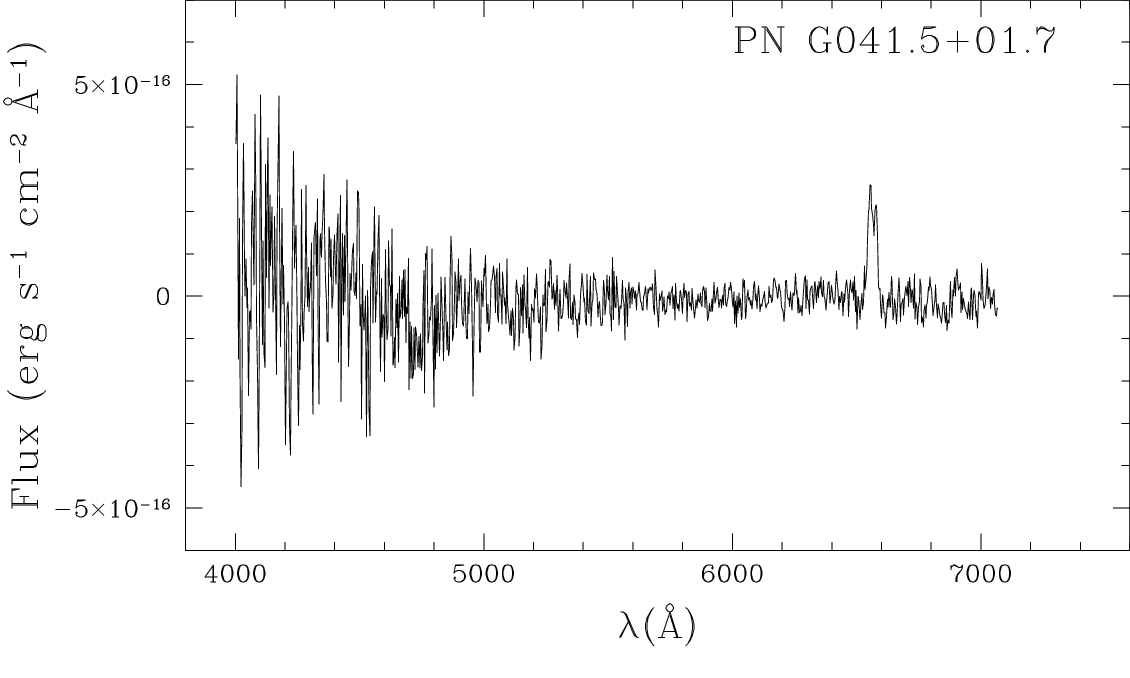}\\
\vspace{0.2 cm}
\includegraphics[scale=0.35]{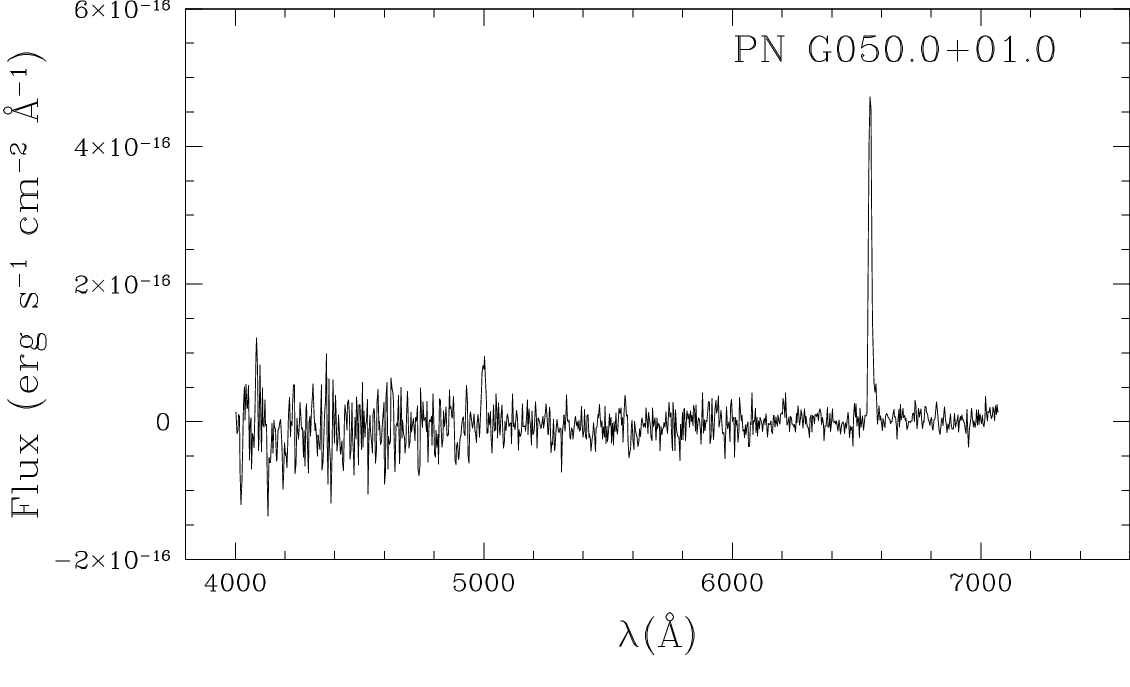} \hspace{0.3 cm} \includegraphics[scale=0.35]{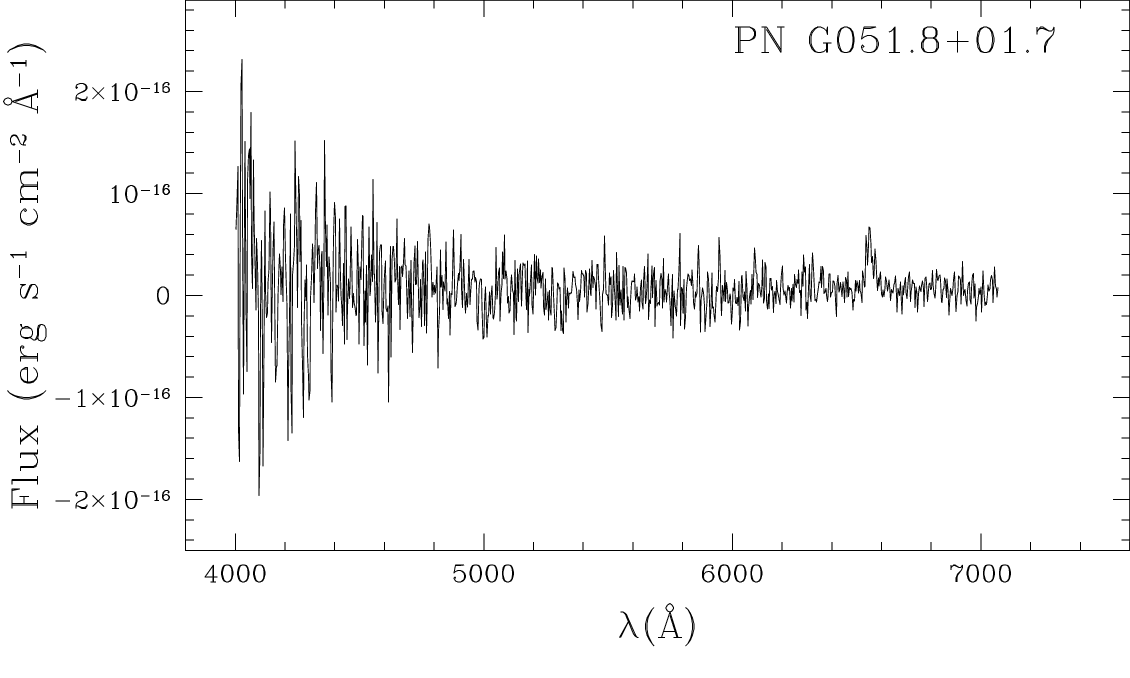}
\end{center}
  \caption{Spectra of 14 objects with a few or no emission lines.}\label{fig:A2}
\end{changemargin}
\end{figure}

\begin{figure}[H]
\addtocounter{figure}{-1}
\begin{changemargin}{-3cm}{-3cm}
\begin{center}
\includegraphics[scale=0.35]{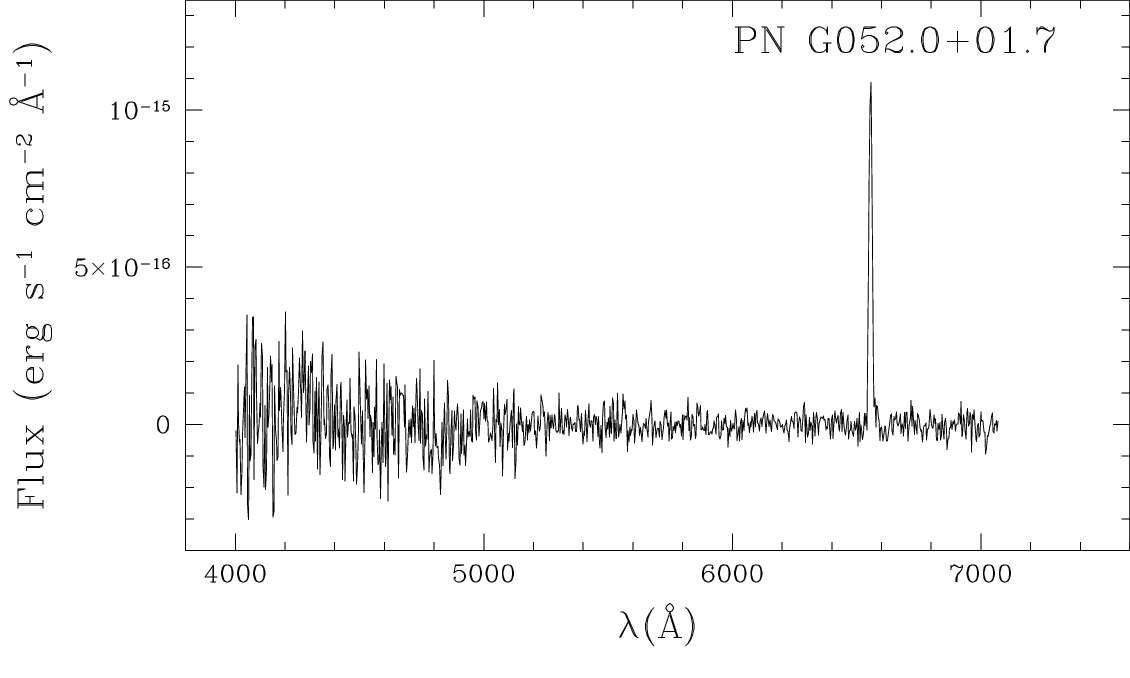} \hspace{0.3 cm} \includegraphics[scale=0.35]{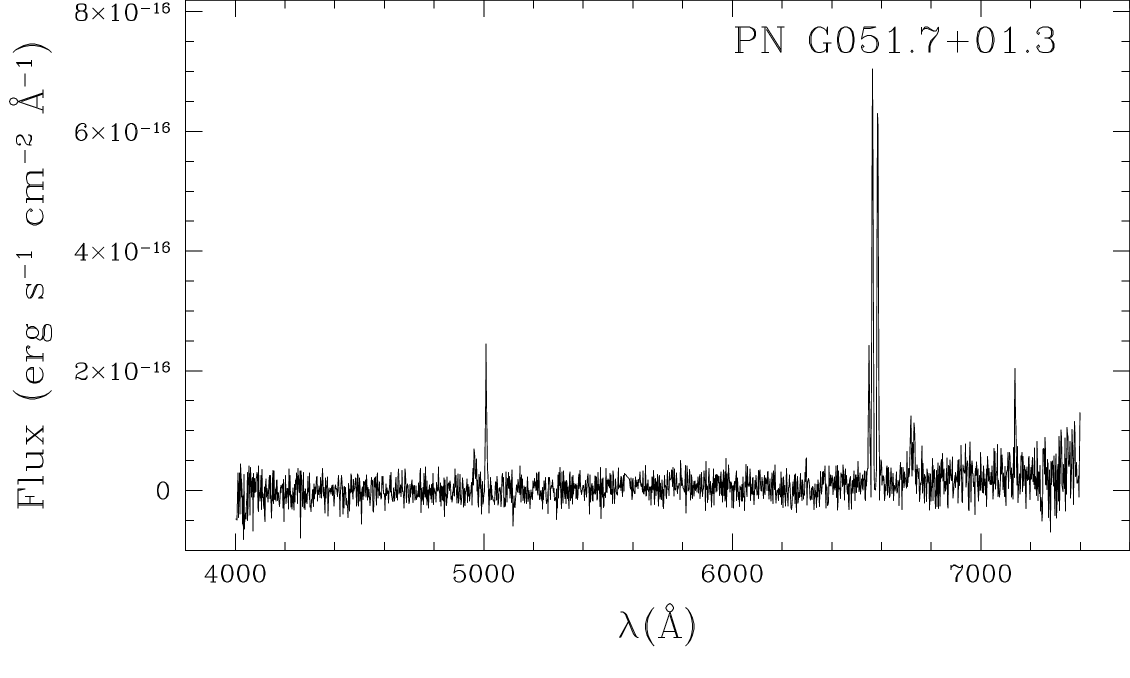}\\
\vspace{0.2 cm}
\includegraphics[scale=0.35]{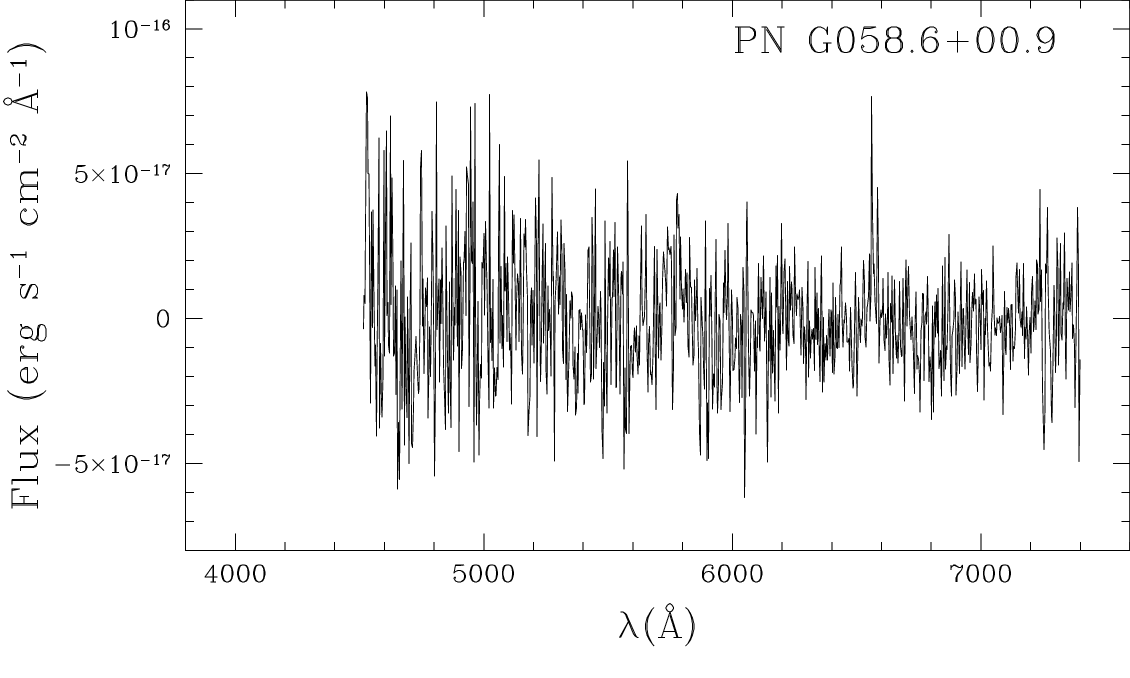} \hspace{0.3 cm} \includegraphics[scale=0.35]{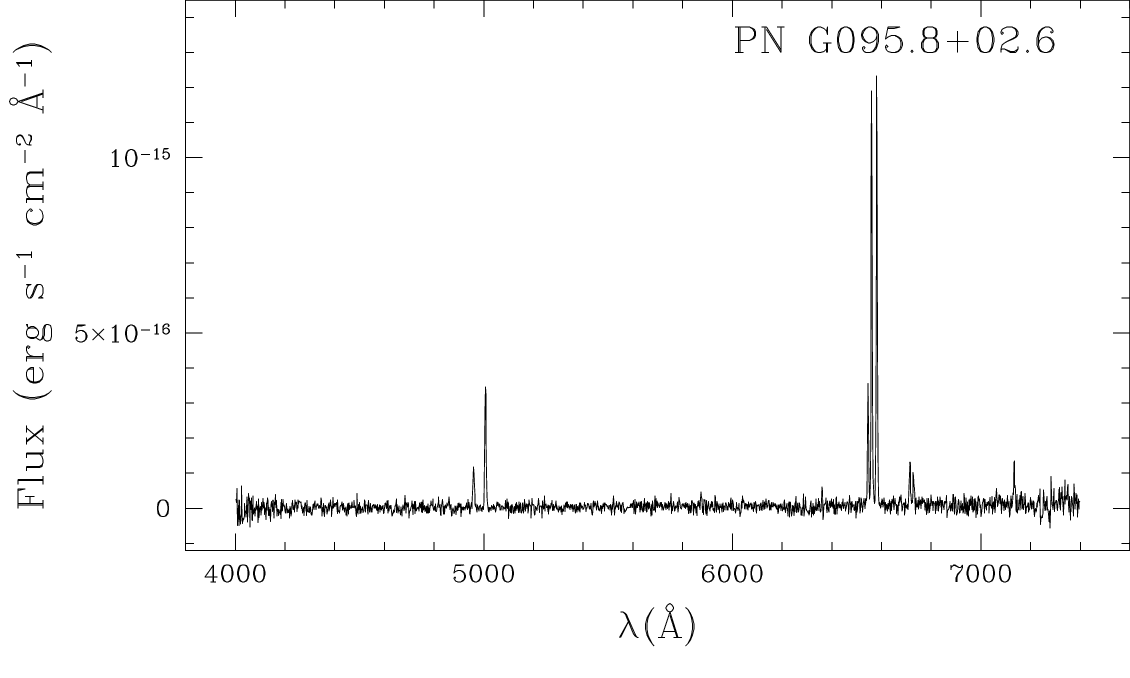}
\end{center}
\caption{(Continued).}
\end{changemargin}
\end{figure}

\newpage

\hypertarget{appendixC}{}
\section*{APPENDIX C: ADDITIONAL LITERATURE INFORMATION} \label{appendix:C}

\renewcommand{\thetable}{A\arabic{table}}

There are different collective works in the literature that involve some of the objects in our sample with topics that are not related to our research. The objects and papers where they appear are listed in Table~\ref{tabApxC}. 

\begin{table}[h!] 
\addtocounter{table}{-8}
\begin{changemargin}{-2cm}{-2cm}
\begin{center} 
\caption{Collective works\lowercase{$^a$} that includes objects in our sample} \label{tabApxC}
\begin{tabular}{lccccccccccc}
\hline 
Name & 1 & 2 & 3 & 4 & 5 & 6 & 7 & 8 & 9 & 10 & 11\\
\hline
PN G003.3+66.1 & ... & ...& ... & $\checkmark$ & ... & $\checkmark$ & ... & ... & ... & $\checkmark$ & $\checkmark$\\
PN G006.5+08.7 & ... & ...& ... & $\checkmark$ & ...& ... & ... & ... & $\checkmark$ & ... & $\checkmark$\\
PN G008.3+09.6 & ... & ...& $\checkmark$ & ... & ...& ... & ... & ... & $\checkmark$ & ... & $\checkmark$\\
PN G031.3+02.0 & ... & ...& ... & ... & ...& ... & ... & ... & ... & ... & $\checkmark$\\
PN G030.5+01.5 & ... & ...& ... & ... & ...& ... & ... & ... & ... & ... & $\checkmark$\\
PN G040.1+03.2 & ... & ...& ... & ... & ...& ... & ... & ... & ... & $\checkmark$ & $\checkmark$\\
PN G038.4+02.2 & ... & ...& ... & ... & ...& ... & ... & ... & ... & $\checkmark$ & ...\\
PN G040.6+01.5 & ... & ...& ... & ... & ...& ... & ... & ... & ... & $\checkmark$ & ...\\
PN G041.5+01.7 & $\checkmark$ & ...& ... & ... & ...& ... & ... & ... & ... & $\checkmark$ & ...\\
PN G047.8+02.4 & ... & ...& ... & ... & ...& ... & ... & ... & ... & ... & $\checkmark$\\
PN G050.0+01.0 & ... & ...& ... & ... & ...& ... & ... & $\checkmark$  & ... & ... & $\checkmark$\\
PN G052.0+01.7 & $\checkmark$ & ...& ... & ... & ...& ... & ... & ... & ... & ... & $\checkmark$\\
PN G051.7+01.3 & ... & ...& ...  & ... & $\checkmark$& ... & ... & ... & ... & $\checkmark$ & ...\\
PN G057.9+01.7 & $\checkmark$ & $\checkmark$ & ... & ... & ...& ... & ... & ... & ... & $\checkmark$ & $\checkmark$\\
PN G058.6+00.9 & $\checkmark$ & ...& ... & ... & ...& ... & ... & $\checkmark$  & ... & ... & $\checkmark$\\
PN G073.6+02.8 & ... & ...& ... & ... & ...& ... & ... & ... & ... & ... & $\checkmark$\\
PN G077.4$-$04.0 & ... & ...& ... & ... & ...& ... & ... & ... & ... & ... & $\checkmark$\\
PN G086.2$-$01.2 & ... & ... & ... & ... & ...& ... & ... & ... & ... & ... & $\checkmark$\\
PN G095.8+02.6 & ... & ...& ... & ... & ...& ... & $\checkmark$ & ... & ... & ... & ...\\
PN G114.4+00.0 & ... & ...& ... & ... & ...& ... & ... & ... & ... & ... & $\checkmark$\\ \hline
\multicolumn{12}{l}{\small$^a$1 \citep{1999ApJS..123..219C} and 3 \citep{2011MNRAS.412..223B} reported flux density at}\\ 
\multicolumn{12}{l}{\small1.4 GHz. 2 \citet{2008ApJ...689..194S} and 6 \citep{2016MNRAS.455.1459F} created a ca-}\\
\multicolumn{12}{l}{\small talog of statistical distances. 4 \citep{2013MNRAS.431....2F} made a catalog of integra-}\\
\multicolumn{12}{l}{\small ted $\mathrm{H{\alpha}}$ fluxes. 5 \citep{2015MNRAS.454.2586F} have identified extended $\mathrm{H_{2}}$ featu-}\\
\multicolumn{12}{l}{\small res. 7 \citep{2017MNRAS.470.3707R} found that the spatial distribution of $\mathrm{H_{2}}$}\\
\multicolumn{12}{l}{\small confirms its bipolar morphology. 8 \citep{2018MNRAS.480.2423I} reported flux at 5}\\ 
\multicolumn{12}{l}{\small GHz and brightness temperature. 9 \citep{2021MNRAS.506.5223J} reported a study}\\
\multicolumn{12}{l}{\small of the possible light curve variability on their CSs. 10 \citep{2021ApJ...919..121H}}\\
\multicolumn{12}{l}{\small made radio observations to find evidence of cold plasma component coexist-}\\ 
\multicolumn{12}{l}{\small ing with hot plasma, but the objects were undetected by the experiment.}\\
\multicolumn{12}{l}{\small 11 \citep{2023ApJS..266...34G} created an UV and optical catalog.}
\end{tabular} 
\end{center}
\end{changemargin}
\end{table}           

Some individual notes for other objects are:\\

\textit{PN G003.3+66.1 (SkAc 1):} Its central star is very well studied by several authors, such as \citet{2016MNRAS.455.3413K} showing the spectra of the white dwarf, \citet{2020A&A...640A..10W} indicating its spectral classification and \citet{2021MNRAS.508.3877G} listing stellar properties. Also, \citet{2015MNRAS.448.3132D} present an image using [\ion{O}{III}] $\lambda5007$ filter. Although it does not appear in the diagnostic diagram we presented in Figure \ref{fig:2}, \citet{2003A&A...405..951M} classified this object as probable PN due to its low heliocentric radial velocity ($\mathrm{-17\pm 2\, km \, s^{-1}}$).\\

\textit{PN G339.4+29.7 (PN Y$-$C 2$-$17):} This object was discovered originally by \citet{1973A&AS...11..335C} and it was described as a PN using only morphological features on a photographic plate; eventually, was cataloged as a Possible PN in \citet{2001A&A...378..843K} and no other studies have been conducted to date. HASH status is Not PN, while SIMBAD status is PN candidate. We have not found any emission lines in our spectrum but we have identified the object as an evolved star of spectral type K3I-II.\\

\textit{PN G358.5+09.1 (Terz N 26, LEDA 89007):} Discovered by \citet{1985Msngr..42....4T} who described it as a PN-like diffuse object from a set of R and B plates. Although \citet{2001A&A...378..843K} cataloged it as Possible PN, some studies in the literature for the last 20 years consider this object a galaxy, e.g.\@, \citet{2003A&A...412...45P} identified it as an elliptical galaxy, \citet{2016A&A...588A..14T} reported a photometric redshift of 0.037633 $\pm$ 0.000167, and a distance of 165.254 Mpc. HASH status is possible galaxy, SIMBAD status is PN candidate, and NED status is galaxy. Our spectrum shows no emission lines or apparent absorption features. Therefore, it is not possible to draw a more definitive conclusion. \\

\textit{PN G358.4+08.9 (Terz N 29, LEDA 89010):} \citet{1985Msngr..42....4T} described it as a red star surrounded by nebulosity as it appeared in R and B plates. There are few studies in the literature, but \citet{2000MNRAS.316..326H} identify this object as a type E galaxy and report a redshift of 0.027786 $\pm$ 0.000167. Nevertheless, \citet{2001A&A...378..843K} cataloged it as Possible PN. HASH status is a possible galaxy, SIMBAD status is PN candidate, and NED status is Galaxy. We have not found emission lines in our spectrum but several absorption features corresponding to G-band, Mgb, NaD and $\mathrm{H{\alpha}}$ typical of elliptical galaxies. We derived a redshift of $\mathrm{z}=0.0279$ that is in complete agreement with previously reported value \citep{2000MNRAS.316..326H}.\\

\textit{PN G058.6+00.9 (PN PM 1$-$309, IRAS 19353+2302):} \citet{1988A&AS...76..317P} proposed this object as a possible PN based on the far-IR colours of the source and reported density fluxes at 12, 25, 60, and 100 $\mu$m. He calculated dust temperature assuming that is the unique source observed and estimated heliocentric distance (5.1 kpc).

\end{appendices}

\end{document}